\documentclass[fleqn,usenatbib]{mnras}

\usepackage[T1]{fontenc}
\usepackage{ae,aecompl}

\usepackage{eso-pic}

\usepackage{graphicx}	
\usepackage{amsmath}	
\usepackage{amssymb}	

\usepackage{multicol}
\usepackage[normalem]{ulem}
\usepackage{bm}

\definecolor{grey}{rgb}{0.75,0.75,0.75}
\definecolor{Orange}{rgb}{1.0,0.5,0.15}
\definecolor{brown}{rgb}{0.7,0.25,0.0}
\definecolor{pink}{rgb}{1.0,0.5,0.5}
\definecolor{darkerred}{rgb}{0.8,0,0}
\definecolor{darkerblue}{rgb}{0,0,0.8}
\definecolor{Blue}{rgb}{0,0.08,0.65}
\definecolor{Red}{rgb}{0.65,0.08,0.05}
\definecolor{Green}{rgb}{0.15,0.45,0.25}

\newcommand{\pd}{\partial}
\newcommand{\rcm}{\bm{r}_{\mathrm{cm}}}

\allowdisplaybreaks

\title[Mean energy overdensity]{Excursion set peaks in energy as a model for haloes}

\author[ Musso \& Sheth]{
  \parbox{\textwidth}{
    Marcello Musso$^{1,2,3,4}$\thanks{E-mail: \texttt{\rm \texttt{mmusso@eaifr.org}}}, and Ravi~K.~Sheth$^{5,6}$\thanks{E-mail: \texttt{\rm \texttt{shethrk@upenn.edu}}}}\\
 \vspace{0.cm}\\~\\
 $^{1}$ ICTP East African Institute for Fundamental Research,
 University of Rwanda, Kigali, Rwanda\\
 $^{2}$ Max-Planck-Institut f\"ur Astrophysik, Karl-Schwarzschild Str. 1, 85741 Garching, Germany\\
 $^{3}$ Deutsches Elektronen-Synchrotron DESY, Notkestr. 85, 22607 Hamburg, Germany\\
 $^{4}$ Institute for Fundamental Physics of the Universe, Via Beirut 2, 34151 Trieste, Italy \\
 $^{5}$ Center for Particle Cosmology, University of Pennsylvania, Philadelphia, PA 19104, USA\\
 $^{6}$ The Abdus Salam International Center for Theoretical Physics, Strada Costiera 11, Trieste 34151 – Italy
}

\pubyear{2021}

\begin{document}
\label{firstpage}
\pagerange{\pageref{firstpage}--\pageref{lastpage}}
\maketitle

\begin{abstract}
  The simplest models of dark matter halo formation rely on the heuristic assumption, motivated by spherical collapse, that virialized haloes originate from initial regions that are maxima of the smoothed matter density field. Here, we replace this notion with the dynamical requirement that protohaloes be regions where the local gravitational flow converges to a point. For this purpose, we look for spheres whose acceleration at the boundary -- relative to their center of mass -- points towards their geometric center: that is, spheres with null dipole moment. We show that these configurations are minima of the energy, corresponding to the most energetically bound spheres.  Therefore, we study peaks of the smoothed energy overdensity field.  This significant conceptual change is technically trivial to implement: to change from density to energy one need only modify the standard top-hat smoothing filter.  However, this comes with the important benefit that, for power spectra of cosmological interest, the model is no longer plagued by divergences: improving the physics mends the mathematics. In addition, the ``excursion set'' requirement that the smoothed matter density crosses a critical value can be naturally replaced by a threshold in energy. Measurements in simulations of haloes more massive than $10^{13}h^{-1}M_\odot$ show very good agreement with a number of generic predictions of our model.
\end{abstract}




\section{Introduction}\label{sec:intro}
Virialized dark matter haloes play an important role in models of nonlinear strucure formation \citep{cs02}.  The most basic quantity of interest is the comoving number density of haloes as a function of halo mass.  Following \cite{ps74}, this halo abundance retains memory, i.e. depends on the statistics, of the primordial fluctuation field.  The simplest models, inspired by spherical collapse, assume that the initial overdensity within a spherical region must equal (exceed) a critical value if it is to become a virialized halo today (before today).   Energy conservation arguments are used to estimate its final density \citep{gg72}.

This assumption has led to extensive study of the abundance of such critically overdense patches in the suitably smoothed initial overdensity fluctuation field.  To avoid double-counting patches which are overlapping with or contained within  larger ones, there are two key approximations:  one is that such patches must be local peaks of the smoothed field \citep{bbks86}, and the other is that they must not be as overdense on larger scales \citep{bcek91}.  These requirements can be combined \citep{jrb89,aj90,manrique1995}, and together they define the excursion set peaks approach of \cite{ps12} and \cite{psd13}, which provides a good description of halo statistics (mass function and bias) once some input from simulations is included to tune the value of the critical overdensity.

However, some problems with the peak based approach remain open. First, when investigating the predictions of this model on a halo by halo basis, at small masses not all protohaloes are peaks of the initial density field. About 25\% of the small haloes are missed by this prescription \citep{ludlowpeaks}. Secondly, for a $\Lambda$CDM power spectrum and a top-hat filter in real space (the most physically motivated smoothing), the actual calculation of the peak abundance contains divergences and cannot be carried out. Thirdly, as we discuss in this paper, a density peak does not coincide with the center of mass of the peak patch. Therefore, the motion relative to the center of mass of the particles around it may not be well approximated by spherical collapse.

From a dynamical point of view, the total energy density enclosed in some reference volume is often a more relevant quantity than the mass density.  If the initial density profile is not flat, then the relation between the two overdensities is stochastic \cite[e.g.][]{bm96}.  As a result, a peak in energy overdensity is not the same as one in matter.  In this paper, we study peaks in the (absolute value of the) enclosed energy density. That is, we look for the most energetically bound regions in the initial density field. 

As we describe below, this approach shows considerable promise for at least three reasons. First, the mean energy overdensity governs the evolution of the moment of inertia of the collapsing patch \cite[e.g.][]{chandra}, which is commonly used to describe virialization. Second, for a sphere, the gradient of the mean energy is proportional to the dipole moment: when it vanishes, the center of mass coincides with the center of the sphere, and the acceleration of particles at the surface points towards it. This is therefore the spherical surface whose actual evolution (described e.g. by perturbation theory) most closely matches spherical collapse. I.e., the position around which the spherical collapse approximation is most accurate. Third, the statistics of the mean energy overdensity have better convergence properties than those of the matter overdensity: one can thus build a self-consistent peak theory without having to tweak the top hat filter as is usually done \citep[e.g.][]{manrique1995,psd13,css17}.  Thus, characterizing protohaloes as peaks of the energy overdensity field addresses the problems of the matter density based excursion set peaks approach.

Section~\ref{sec:phys} discusses the physical motivation for working with energy rather than density as the primary variable.  After first setting up notation and highlighting a number of interesting relations between derivatives of energy and overdensity, Section~\ref{sec:stats} derives our expression for the comoving number density of excursion set peaks in energy and describes a number of generic features of the approach.  Section~\ref{sec:sims} compares a number of these predictions with the properties of protohaloes identified in simulations.  A final section summarizes some consequences of our analysis and discusses a number of interesting directions for further study.  Appendix \ref{app:acc} contains some technicalities of the multipole expansion, Appendix~\ref{app:corrs} provides expressions for some of the important correlations between matter and energy overdensities, and Appendix~\ref{app:alt} discusses a modification of our energy peaks approach:  a hybrid model which uses both energy and density to identify protohaloes.

\section{Matter, energy and motion}\label{sec:phys}

\subsection{Matter vs energy overdensity}
\label{sec:energy}

Calling $\rho(\bm{r})$ the matter density at $\bm{r}$ and $\bar\rho$ its background value, the mean \emph{matter} overdensity within a sphere of physical radius $R$ centered at the origin is usually defined as 
\begin{equation}
  \delta_R \equiv
  \int_{V_R} \frac{{\rm d}{\bm r}}{V_R}\,\delta({\bm r})\,,
  \label{eq:d}
\end{equation}
where $\delta(\mathbf{r})\equiv\rho(\mathbf{r})/\bar\rho-1$  and $V_R\equiv 4\pi R^3/3$.  The potential energy in the sphere due to matter is \cite[e.g.][]{chandra,BinneyTremaineBook1987} 
\begin{equation}
  U =  \int_{V_R}\mathrm{d}\bm{r}\,\rho(\bm{r})
  \,(\bm{r}-\bm{r}_{\mathrm{cm}}) \cdot\bm{g}\,,
\end{equation}
where $\rcm$ is the center of mass position, defined as
\begin{equation}
  \rcm \equiv \frac{1}{M}\!\int_{V_R}\!
  \mathrm{d}\bm{r}\,\rho(\bm{r})\,\bm{r} =
  \frac{\bar\rho}{M}\!\int_{V_R}\!
  \mathrm{d}\bm{r}\,\delta(\bm{r})\bm{r}\,,
\label{eq:rcm}
\end{equation}
where $M$ is the total mass in $V_R$, and $\bm{g}\equiv -\nabla\Phi + [\nabla\Phi]_{\mathrm{cm}}$ is the acceleration at $\bm{r}$, induced by matter only, relative to the acceleration of the center of mass.

The acceleration splits into background and peculiar: 
$\nabla\Phi = 4\pi G\bar\rho(\bm{r}/3+\nabla\phi)$, with the potential perturbation normalized so that $\nabla^2\phi =\delta$ in physical coordinates. The relative acceleration becomes 
\begin{equation}
  \bm{g} = 
  - 4\pi G\bar\rho \bigg[\frac{\bm{r}- \bm{r}_{\mathrm{cm}}}{3}
  + \nabla\phi(\bm{r})
  - \int_{V_R}\frac{\mathrm{d}\bm{r}'\rho(\bm{r}')}{M}\nabla\phi(\bm{r}')
  \bigg],
\label{eq:g}
\end{equation}
the last term in the bracket being the peculiar acceleration of the center of mass $-[\nabla\phi]_{\mathrm{cm}}$.
For comparison, the acceleration from the cosmological constant is simply $(\Lambda/3)(\bm{r}-\bm{r}_{\mathrm{cm}})$.
The potential energy $U$ can then be written as
\begin{equation}
  U = -4\pi G \bar\rho
  \frac{MR_I^2}{5} (1+\epsilon_R)\,,
\end{equation}
where 
\begin{equation}
  R_I^2 \equiv
  \frac{5}{3M}\int_{V_R} \!\mathrm{d}\bm{r}\,\rho(\bm{r})\,
  |\bm{r}-\bm{r}_{\mathrm{cm}}|^2\,  \equiv \frac{5}{3M} T
\label{eq:RI}
\end{equation}
is the (square of the) inertial radius, and
\begin{equation}
  \epsilon_R =
  \frac{5}{MR_I^2}\int_{V_R}\!\mathrm{d}\bm{r}\,\rho(\bm{r})
  \,(\bm{r}-\rcm)\cdot\nabla{\phi(\bm{r})}
\label{eq:eps}
\end{equation}
is the potential \emph{energy} overdensity associated with $V_R$.
It plays for $U$ the same role that the matter overdensity $\delta_R$ plays for the mass $M=\bar\rho V_R(1+\delta_R)$. The contribution from the peculiar acceleration of the center of mass is constant over the sphere and so drops out of equation \eqref{eq:eps}.

At leading order in perturbations, in equation \eqref{eq:eps} one can replace $\rho(\bm{r})/M\simeq 1/V_R$ and, for a spherical volume, $R_I\simeq R$ and $\rcm\simeq0$. We can thus redefine the \emph{linearized} mean energy overdensity as
\begin{equation}
  \epsilon_R\equiv 
  \frac{15}{4\pi R^5}\int_{V_R}\mathrm{d}\bm{r}\,
  \bm{r}\cdot\nabla{\phi(\bm{r})}\,.
\label{eq:Elin}
\end{equation}
If $V_R$ is a sphere, only the radially directed monopole term of the peculiar gravitational acceleration, $\nabla\phi \sim (\delta_r/3)\,\bm{r}$, survives after integrating over the angles, and thus
\begin{equation}
  \epsilon_R = \frac{5}{R^5}\int_0^R {\rm d}r\,r^4\,\delta_r
\end{equation}
\cite[e.g.][equation 2.27b]{bm96}. For a homogeneous sphere,  $\delta({\bm r})$ is constant, $R_I=R$ and $\delta_R = \epsilon_R$. This returns the familiar result $U = -(5/3)GM^2/R$. In general, however, $\delta_R\ne \epsilon_R$.  Note also that, although we defined $\delta_R$ and $\epsilon_R$ using physical coordinates, they have exactly the same expression in comoving coordinates.

At a generic position ${\bm x}$, and for an arbitrary configuration of the density field, the matter overdensity in a spherical volume $V_R$ is a weighted sum over the Fourier modes:
\begin{equation}
  \delta_R({\bm x}) 
  = \int \frac{{\rm d}{\bm k}}{(2\pi)^3} \,\delta({\bm k})\,
  W_1(kR)\,{\rm e}^{i{\bm k\cdot \bm x}},
 \label{eq:Wth}
\end{equation}
where $W_1(z) \equiv 3j_1(z)/z$. The expression for the mean linear energy overdensity at the same position follows inserting the Fourier modes $(-i\bm{k}/k^2) \delta(\bm{k}){\rm e}^{i{\bm k\cdot \bm r}}$ of the peculiar acceleration $\nabla\phi$ in equation \eqref{eq:Elin}; integrating over $\bm{r}$ gives
\begin{equation}
  {\epsilon}_R({\bm x}) 
  = \int \frac{{\rm d}{\bm k}}{(2\pi)^3} \,\delta({\bm k})\, W_{2}(kR)
   \,{\rm e}^{i{\bm k\cdot \bm x}},
 \label{eq:We}
\end{equation}
where $W_{2}(z) \equiv 15j_2(z)/z^2$. Replacing $\delta_R$ with $\epsilon_R$ simply corresponds to a change of filter. At $z\gg 1$ this filter decays faster than $W_{1}$ by one power of $z$.  

As we noted in the introduction, an approach based on peaks in $\delta_R$ leads to divergences. Therefore, it is interesting to explore what happens if one studies peaks in $\epsilon_R$ instead. The more convergent asymptotic behaviour of $W_2$ compared to $W_1$ suggests that they will be mathematically better behaved. In the next subsections we argue that peaks in $\epsilon_R$ are also physically more reasonable than peaks in $\delta_R$.

\subsection{Haloes as energy overdensity peaks}
Any initial region is displaced and deformed by gravity. Its evolution separates as center of mass displacement plus collapse onto the center of mass.  Only the latter creates a high density region, and it is this process that the spherical collapse model approximates.  In this subsection, we discuss how to choose  locations where this approximation is optimal.

Let us define the dimensionless dipole moment of the sphere $\bm{D}_R \equiv\rcm/R$. From equation \eqref{eq:rcm}, its Fourier expression at leading order (at $\bm{x}=0$ for convenience) is 
\begin{equation}
  \bm{D}_R = \frac{3}{R}\int \frac{{\rm d}{\bm k}}{(2\pi)^3} \,
  \frac{i{\bm k}}{k^2}\, \delta({\bm k})\,j_2(kR)\,.
\label{eq:dipole}
\end{equation}
Having $\bm{D}_R\ne 0$ means that the sphere's center of mass, towards which the  acceleration converges, does not coincide with the geometric center. Hence, the collapse of the object is not isotropic, but one side collapses faster than the other.  To avoid this, and find the locus of the particles that would reach the center at the same time (neglecting higher multipoles), one must shift the sphere until $\bm{D}_R=0$.

One can see this directly by looking at the acceleration $\bm{g}$ of a particle at the surface of the sphere relative to the center of mass, given by equation \eqref{eq:g} with $\rho(\bm{r})/M$ replaced by $1/V_R$ at leading order.
As we show in Appendix \ref{app:acc}, this relative acceleration at $\hat {\bm r}R$ can be written as
\begin{equation}
  g_i = -\frac{GM}{R^2}\bigg[\hat r_i
  + D_{R,j} (3\hat r_i\hat r_j -\delta_{ij})
  +\dots \bigg] 
\label{eq:gmult}  
\end{equation}
using a  multipole expansion. This expression contains: \\[1mm]
(i) ~a monopole term that we treat with spherical collapse; \\
(ii) ~one anisotropic dipole term, which we ask to vanish; \\ 
(iii) ~a double infinite series of higher order multipoles, which we suppose neglible in this work.\\[1mm]
This expansion has only one dipolar anisotropy (since the center of mass acceleration was subtracted off), which can be set to zero by a translation. Dealing with higher order anisotropies would require a different transformation, like a deformation of the spherical boundary.

Combining the $k^2$ and $j_2$ factors in equation~\eqref{eq:dipole} gives  
\begin{equation}
  \bm{D}_R = \frac{R}{5}
   \int \frac{{\rm d}{\bm k}}{(2\pi)^3} \,i{\bm k}\,\delta({\bm k})\,W_2(kR)\,
    = \frac{R}{5}\,{\bm \nabla}\epsilon_R\,.
\label{eq:dipgrad}  
\end{equation}
The final equality indicates that a sphere with null dipole moment must also have $\bm{\nabla}\epsilon_R=0$:  a convergence point of the local gravitational flow must be a stationary point in the energy overdensity field $\epsilon_R$. Moreover,
\begin{equation}
  \nabla_i\nabla_j\epsilon_R
  = \frac{5}{R}
  \int \frac{{\rm d}{\bm k}}{(2\pi)^3} \,\frac{k_ik_j}{k^2}\,
  \delta({\bm k})\, \frac{\mathrm{d} W_1(kR)}{\mathrm{d} R}\,,
\end{equation} 
and in particular $\nabla^2\epsilon_R=(5/R)\mathrm{d}\delta_R/\mathrm{d} R$, which describes the change in infall velocity as $R$ grows.  For it to be slower from any direction, so that larger (outer) shells collapse later than inner ones, the Hessian $\nabla_i\nabla_j\epsilon_R$ must be negative definite. Thus, we seek stationary points of $\epsilon_R$ that are peaks.

\subsection{The collapse of the inertial radius}

The total energy contained inside the sphere, normalized like in equation \eqref{eq:RI} i.e. in physical coordinates, is
\begin{equation}
  E \equiv \frac{5}{3M}\bigg(\!K+U -\frac{\Lambda}{6}T\bigg),
\end{equation}
where $K\equiv\int_{V_R}\mathrm{d}\bm{r}\rho(\bm{r})|\dot{\bm r}-\dot{\bm r}_{\mathrm{cm}}|^2/2$ is the kinetic energy and $T$ is the moment of inertia defined in equation \eqref{eq:RI}. It can be conveniently written as
\begin{equation}
  E = \frac{\dot R_I^2}{2} + kR_I^2 - \frac{GM_I}{R_I}
  - \frac{\Lambda}{6}R_I^2\,,
\label{eq:E}
\end{equation}
where $k\equiv (5/6M)\!\int_{V_R}\!\!\mathrm{d}\bm{r}\rho(\bm{r})v_I^2$, with $\bm{v}_I\equiv[(\bm{r}-\rcm)/R_I]\,\dot{}$, is the peculiar kinetic energy with respect to $R_I$ as a local scale factor, and $M_I\equiv 5UR_I/3GM = (4\pi/3)\bar\rho(1+\epsilon_R)R_I^3$ is an effective mass.  The true mass is $M=(4\pi/3)\bar\rho(1+\delta_R)R^3$. Since $\delta_R\ll 1$ initially, $M\simeq (4\pi/3)\bar\rho R^3$.

Zel'dovich initial conditions, relating peculiar velocities to the potential perturbation, give $\dot{\bm r}\simeq H(\bm{r}+\nabla\phi)$ in an EdS universe. When inserting this in $K$, assuming zero curvature, the background contributions to $E$ cancel, leaving
\begin{equation}
  E\simeq -\frac{GM_I}{R_I}\epsilon_R\,.
\label{eq:Ein}
\end{equation}
In a $\Lambda$CDM cosmology, a factor of $(3/5 + 2f/5\Omega_m)$ multiplies this expression, where $f=\mathrm{d}\ln D/\mathrm{d}\ln a$ and $D$ is the growth function of linear matter perturbations ($f=1$ in EdS). For an isolated object, or for a portion of a spherically symmetric density profile, this energy is conserved.

In terms of $R_I$, the virial equation for the evolution of the moment of inertia, $\ddot T/2 = 2K + U + \Lambda T/3$, is 
\begin{equation}
  \ddot R_I = 2kR_I-\frac{GM_I}{R_I^2}+\frac{\Lambda}{3}R_I
\label{eq:RIddot}
\end{equation}
Since $k$ is of second order in perturbations, and at first order $M_I$ and $E$ are conserved\footnote{Assuming $\dot M_I\simeq0$ is self-consistent, since a spherical collapse solution has $R_I\simeq a\,(1-\epsilon_R/3)$ at linear order. It also follows from $\dot U = -\int_{V_R}\!\mathrm{d}\bm{r}\,\rho(\bm{r})\,\dot{\bm r}\cdot\bm{g}$ (note the sign!) for an isolated body \citep{chandra,BinneyTremaineBook1987}, so that $\dot M_I\propto -\int_{V_R}\!\mathrm{d}\bm{r}\,\rho(\bm{r})\,\bm {v}_I\cdot\bm{g}\simeq0$ ar first order. The coupling of a sphere with its environment appears only at second order.}, equations \eqref{eq:E}, \eqref{eq:Ein} and \eqref{eq:RIddot} coincide (up to first order) with the equations of spherical collapse for $R$, except that one replaces $\delta_R\to\epsilon_R$. Thus, $R_I$ approximately follows a spherical collapse solution for overdensity $\epsilon_R$, with corrections starting at second order in perturbation theory.
Hence, $\epsilon_R$ plays for the inertial radius $R_I$ the role that $\delta_R$ plays for $R$: it indicates if and how fast the moment of inertia of an extended region is destined to shrink. Therefore, $\epsilon_R$ may be at least as good an indicator as $\delta_R$ for inferring if an initial patch is destined to become a nonlinear overdense halo.
Although we remarked earlier that $\delta_R\ne \epsilon_R$ in general, we expect they will be correlated with one another.  We quantify this and a number of other relevant correlations, in the next section.

\section{Excursion set peaks of the mean energy overdensity}\label{sec:stats}
Having laid out the physical motivation for studying peaks in $\epsilon_R$, we now discuss their statistics.

\subsection{Normalized variables}
\label{sec:norm}
\begin{figure}
  \includegraphics[width=\columnwidth]{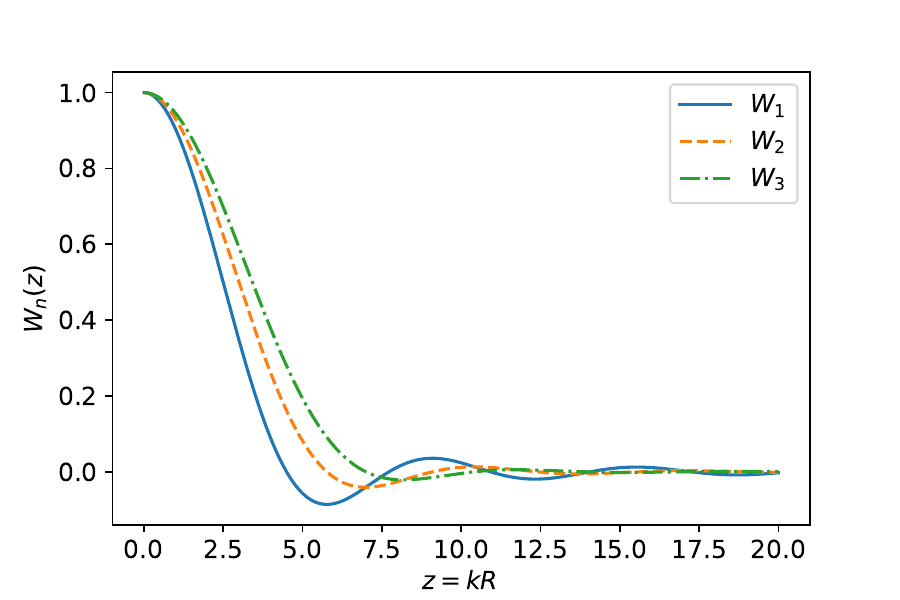}
\caption{\label{fig:filters} Plot of the filters $W_n(kR)$ for $n=1,2,3$, showing faster convergence for higher values of $n$. Since $W_2$ converges faster than $W_1$, it is possible to compute peak statistics of the $\epsilon_R(\mathbf{x})$ field with a $\Lambda$CDM power spectrum without incurring divergences.}
\end{figure}

In this subsection we lay out the notation that will be needed in what follows.
We begin by defining the generic filter
\begin{equation}
  W_n(z) \equiv (2n+1)!!\, \frac{j_n(z)}{z^n}\,,
\end{equation}
such that $W_n(z)\to1$ as $z\to0$.  
The recursion relations of the spherical Bessel functions imply that 
\begin{align}
  &\mathrm{d}W_n(z)/\mathrm{d}\ln z = -z^2W_{n+1}(z)/(2n+3)\,, 
  \label{eq:recursion} \\
  &W_n(z) - W_{n-1}(z) = z^2W_{n+1}(z)/[(2n+3)(2n+1)]\,,
  \label{eq:rec2}  
\end{align}
which we use below.
We also define the variances
\begin{equation}
 \sigma_{jn}^2(R) \equiv \int \frac{{\rm d}k}{k}\,k^{2j}\,\frac{k^3 P(k)}{2\pi^2}\,W^2_{n}(kR)\,.
 \label{eq:sigman}
\end{equation}
Because $W_2$ decays faster than $W_1$ at large $k$, $\sigma_{j2}^2$ may converge even when $\sigma_{j1}^2$ does not.  We will see that, as a result, peaks theory for $\epsilon_R$ is better behaved than for $\delta_R$.
The exact behavior of the different filters is shown in Fig.~\ref{fig:filters}. Setting $z=kR$ and Fourier transforming $W_n(kR)$ yields 
\begin{equation} 
  {\cal W}_n(r) = \begin{cases}
            \frac{\Gamma(n+3/2)}{\Gamma(n)\Gamma(1/2)}\frac{[1 - (r/R)^2]^{n-1}}{\pi R^3} &\mbox {if}\quad  r\le R \\
                       0                      &\mbox{otherwise}
           \end{cases}
  \quad .
\end{equation}
Evidently, in real space, filters with larger $n$ weight the central regions more than the outer parts.

Secondly, we introduce the normalized fields
\begin{equation}
  \nu_R(\bm{x}) \equiv
  \frac{\delta_R(\bm{x})}{\sigma_{01}(R)} \qquad{\rm and}\qquad
  \omega_R(\bm{x}) \equiv
  \frac{\epsilon_R(\bm{x})}{\sigma_{02}(R)} \,.
\end{equation}
Next, we need the (normalized) gradient
\begin{equation}
  \eta_{i,R} \equiv \frac{\nabla_i\epsilon_R}{\sigma_{12}(R)}
  = \int \frac{{\rm d}{\bm k}}{(2\pi)^3} \,\frac{ik_i\,\delta({\bm k})}{\sigma_{12}(R)}\,W_2(kR) \,{\rm e}^{i{\bm k\cdot \bm x}}
 \label{eq:etaR}
\end{equation}
and the (negative of the) normalized Hessian
\begin{equation}
  \zeta_{ij,R} \equiv -\frac{\nabla_i\nabla_j\epsilon_R}{\sigma_{22}(R)}
  = \int \frac{{\rm d}{\bm k}}{(2\pi)^3} \,\frac{k_ik_j\,\delta({\bm k})}{\sigma_{22}(R)}\, W_2(kR)\,{\rm e}^{i{\bm k\cdot \bm x}},
 \label{eq:zetaR}
\end{equation} 
where $\sigma_{12}^2(R) = \langle {\bm \nabla} \epsilon_R\cdot {\bm\nabla}\epsilon_R\rangle$ and
$\sigma_{22}^2(R) = \langle[\nabla^2 \epsilon_R]^2\rangle$
were defined in equation \eqref{eq:sigman}. These definitions differ from what is commonly used in the literature \cite[e.g.][]{bbks86} only because we use the filter $W_2$.

We then define the curvature $x_R \equiv {\rm Tr}(\zeta_R)$ as the trace of the (negative of the) normalized Hessian: 
\begin{equation}
  x_R\equiv  -\frac{\nabla^2 \epsilon_R}{\sigma_{22}(R)}
  = \int \frac{{\rm d}{\bm k}}{(2\pi)^3} \,
  \frac{k^2\,\delta({\bm k})}{\sigma_{22}(R)}\, W_2(kR)\,
  {\rm e}^{i{\bm k\cdot \bm x}}\,,
 \label{eq:slopeR}
\end{equation}
normalized so that $\langle x_R^2\rangle=1$. 
Finally, we introduce the (normalized) `slope' of the mean energy overdensity
\begin{equation}
  y_R \equiv -
  \frac{7\,\mathrm{d}\epsilon_R/\mathrm{d}\ln R}{R^2\sigma_{23}(R)}
  = \int \frac{{\rm d}{\bm k}}{(2\pi)^3} \frac{k^2\,\delta({\bm k})}{\sigma_{23}(R)}\, W_3(kR)\,{\rm e}^{i{\bm k\cdot \bm x}}.
\label{eq:slopeEps}
\end{equation}
The last equality follows from case $n=2$ of equation~(\ref{eq:recursion}).

For what is to follow, it is useful to note that the curvature $x_R$ can also be written as
\begin{equation}
  x_R  = -\frac{5\,{\rm d}\delta_R/{\rm d} \ln R}{R^2\sigma_{22}(R)}\,,
\label{eq:x-slope}
\end{equation}
(equation~\ref{eq:recursion} with $n=1$).  The derivative $-{\rm d}\delta_R/{\rm d}\ln R$ on the right hand side is sometimes called the `slope' of the mean matter overdensity.  In the standard analysis, the `slope' and `curvature' $-\nabla^2\delta_R/\sigma_{21}(R)$ of the overdensity are the same for Gaussian smoothing filters.  The expression above shows that the slope of the tophat smoothed overdensity is the same as the curvature of the mean energy overdensity.  

This is significant because the slope of the mean matter overdensity plays an important role in the `upcrossing approximation' of the traditional excursion set approach \cite[][although their `slope' variable is really ${\rm d}\delta_R/{\rm d}\sigma_{01}^2$]{ms12}.  The simplest versions of this approach require ${\rm d}\delta_R/{\rm d} \ln R\le 0$ as this implements the request that the outer shells collapse after the inner ones.  Equation~(\ref{eq:x-slope}) shows that by constraining this slope, upcrossing constrains the curvature of the mean energy overdensity $x_R$.

We had argued that we are interested in stationary points of $\epsilon_R$:  positions where ${\bm\nabla}\epsilon_R=0$ (c.f. equation~\ref{eq:dipgrad}).  The `upcrossing' constraint $\mathrm{d}\delta_R/\mathrm{d} \ln R<0$ of equation~\eqref{eq:x-slope} adds the additional requirement that $\mathrm{Tr(\zeta_R)}>0$.  Since the same argument (about outer shells collapsing later than the inner ones) actually applies to the full $\zeta_{ij}$, not just to its trace, $\nabla_i\nabla_j\epsilon_R\propto -\zeta_{ij} $ must be negative definite.  Thus, we are interested in those stationary points of $\epsilon_R$ that are peaks.

Furthermore, for the `slope' of the energy one has that
\begin{equation}
  y_R = 35\frac{\epsilon_R - \delta_R}{R^2\sigma_{23}(R)}\,.
\label{eq:yepsdel}
\end{equation}
Evidently,  requiring ${\rm d}\epsilon_R/{\rm d}\ln R\le 0$ imposes that $\delta_R\leq\epsilon_R$. Physically, this constraint guarantees that the inertial radius $R_I$, which is initially equal to $R$, gradually becomes smaller than $R$ as the density profile develops a denser core.
We exploit this in the next subsection.

In what follows, we will be interested in whether or not $\epsilon_R$ exceeds a critical value.  With some abuse of notation, we use $\epsilon_c(R)$ to denote the fact that this critical value may be different for each $R$.  The associated normalized height and slope of this critical value are given by 
\begin{equation}
  \omega_c \equiv\frac{\epsilon_c}{\sigma_{02}}  \qquad{\rm and}\qquad
  y_c \equiv - \frac{7\,\mathrm{d}\epsilon_c/\mathrm{d}\ln R}{R^2\sigma_{23}}  .
 \label{eq:wcyc}
\end{equation}

\subsection{Peaks in the $\epsilon_R(x)$ field}
Since $\epsilon_R(\bm{x})$ is a Gaussian field, the comoving number density of peaks of height $\epsilon_c$ at fixed comoving scale $R$ can be computed following the steps laid out by \cite{bbks86} for $\delta_R$ \cite[see also][for a shorter derivation based on rotational invariants]{genLagbias}.
Neglecting for simplicity the subscript $R$ on variables, the result is formally written as
\begin{equation}
  \frac{{\rm d}n}{{\rm d}\omega_c} =
  \frac{p(\omega_c)}{V_*}                                     
  \int_0^\infty \!\! {\rm d}x\,f(x)\,p(x|\omega_c)\,,
 \label{eq:dnBBKS}
\end{equation}
where $V_*\equiv [6\pi (\sigma_{12}/\sigma_{22})^2]^{3/2}$ defines a characteristic volume, $p(x|\omega_c)$ is the conditional distribution of $x$ given $\omega=\omega_c$, and 
\begin{align}
  f(x)
  &\equiv \sqrt{\frac{2}{5\pi}}
  \left[\left(\frac{x^2}{2}-\frac{8}{5}\right)e^{-5x^2\!/2}
  +\left(\frac{31}{4}x^2+\frac{8}{5}\right) e^{-5x^2\!/8}\right] \notag \\
  & +\frac{1}{2}\left(x^3-3x\right) 
    \left[\mathrm{erf}\bigg(\sqrt{\frac{5}{2}}x\bigg)
    +\mathrm{erf}\bigg(\sqrt{\frac{5}{8}}x\bigg)\right],
\end{align}
as given by equation~(A15) of \cite{bbks86}, follows from integrating over the traceless part of $\zeta_{ij}$. Therefore, $[f(x)/V_*]\,p(x,\omega_c)$ is the number density of peaks of height $\omega_c$ and curvature $x$.  We note in passing that, had we not integrated over the traceless part of $\zeta_{ij}$, we would have expressions for peak `ellipticity' and `prolateness', exactly as in Appendix~C of \cite{bbks86}, but with our $W_2$ smoothing window.

Bayes' rule can be used to show that $p(x|\omega_c)$ is the Gaussian distribution 
\begin{equation}
  p(x|\omega_c) = \frac{1}{\sqrt{(2\pi)(1-\gamma_{x\omega}^2)}}
  \exp\bigg[-\frac{(x-\gamma_{x\omega}\omega_c)^2}{2(1-\gamma_{x\omega}^2)}\bigg] \,,
  \label{eq:px|ec}
\end{equation}
where the cross-correlation coefficient $\gamma_{x\omega}$ is
\begin{equation}
  \gamma_{x\omega}(R) \equiv \langle x\omega\rangle =
  \int \frac{{\rm d}k}{k}\,\frac{k^3 P(k)}{2\pi^2}\,
  \frac{k^2\,W_2^2(kR)}{\sigma_{22}(R)\sigma_{02}(R)}\,.
\end{equation}
Equation~(\ref{eq:dnBBKS}) describes the differential distribution of the number density of energy peaks of height $\epsilon_c$ when the smoothing scale is $R$. However, it does not say how $R$ must be chosen. Since a given $R$ contains mass $\propto 4\pi R^3/3$, ${\rm d}n/{\rm d}\epsilon_c$ for fixed $R$ is not obviously related to a distribution of halo masses.  To remedy this shortcoming of peaks theory, we must combine it with the excursion set approach.

\subsection{Excursion sets in $\epsilon_R$}
To set the smoothing scale, we add the `excursion set' requirement  that at a fixed position $\bm{x}$ the smoothed field $\epsilon_R$ equals the critical value $\epsilon_c$ on scale $R$, but is smaller on the next larger smoothing scale \cite[the `upcrossing' approximation of][]{ms12}.  As  noted in Section \ref{sec:energy}, this constraint on the slope $y_R$ ensures that the inertial radius $R_I$ shrinks relative to $R$ as the object collapses and its density profile steepens and peaks around its center of mass.  As $\epsilon_c$ may depend on $R$ in general, the upcrossing condition is $\epsilon_R\ge \epsilon_c(R)$ but $\epsilon_{R+\Delta R} \le \epsilon_c(R+\Delta R)$. To first order in $\Delta R$, and using the normalized slope from equation \eqref{eq:slopeEps}, this means $\omega_c\leq\omega_R\leq\omega_c+(R\sigma_{23}/7\sigma_{02})(y-y_c)\Delta R$ and $y\geq y_c$, where $y_c$ is given by equation~(\ref{eq:wcyc}).

Integrating over the allowed ranges and dividing by $\Delta R$ gives the distribution of $R$ values at which $\epsilon_R$ upcrosses:
\begin{equation}
  f_{\mathrm{up}}(R) =
  \frac{R\sigma_{23}}{7\sigma_{02}}\, p(\omega_c)
  \int_{y_c}^\infty \!\! {\rm d}y\,(y-y_c)
  \,p(y|\omega_c)\,,
\label{eq:ES}
\end{equation}
where
\begin{equation}
  p(y|\omega_c) =
  \frac{1}{\sqrt{(2\pi)(1-\gamma_{y\omega}^2)}}
  \exp\bigg[-\frac{(y-\gamma_{y\omega}\omega_c)^2}{2(1-\gamma_{y\omega}^2)}\bigg]
\label{eq:pomega|ec}
\end{equation}
is the Gaussian conditional distribution for $y$, with
\begin{equation}
  \gamma_{y\omega}(R) \equiv \langle y\omega\rangle =
  \int \frac{{\rm d}k}{k}\,\frac{k^3 P(k)}{2\pi^2}\,
  \frac{k^2\,W_3(kR)W_2(kR)}{\sigma_{23}(R)\sigma_{02}(R)}\,.
\end{equation}
This result closely resembles that of equation \eqref{eq:dnBBKS}, even more so when $y_c=0$ (for constant $\epsilon_c$).

Using $Rf_{\mathrm{up}}(R) = \sigma_{02}\,f_{\mathrm{up}}(\sigma_{02})\,|{\rm d}\ln\sigma_{02}/{\rm d}\ln R|$ and integrating over $y$ gives
\begin{equation}
  f_{\mathrm{up}}(\sigma_{02}) =   \frac{\omega_*}{\sigma_{02}}
  \frac{{\rm e}^{-\omega_c^2/2}}{\sqrt{2\pi}}
  F\bigg(\frac{\gamma_{y\omega}\omega_*}{\sqrt{1-\gamma_{y\omega}^2}}\bigg)\,,
\label{eq:fup}
\end{equation}
where
\begin{equation}
  \omega_* \equiv \omega_c-y_c/\gamma_{y\omega}
  = -{\rm d}\omega_c/{\rm d}\ln\sigma_{02}
 \label{eq:w*}
\end{equation}
reduces to $\omega_c$ for a constant barrier, and
\begin{equation}
  F(z) \equiv \frac{1 + \mathrm{erf}(z/\sqrt{2})}{2}
  + \frac{\mathrm{e}^{-z^2/2}}{\sqrt{2\pi}z}\,.
\label{eq:F}
\end{equation}
Equation \eqref{eq:fup} describes the distribution of smoothing radii $R$ satisfying the upcrossing constraint when the smoothing filters are centered on randomly chosen positions in the field.  In particular, these positions need not be peaks.

\subsection{Excursion set peaks of $\epsilon_R$}

The excursion set peaks approach \citep{ps12,psd13} posits that the comoving number density of protohalo patches of initial size $R$ (that will become haloes of mass $M=\bar\rho\,4\pi R^3/3$) can be estimated by combining the excursion sets and peaks analyses of the previous two subsections.  I.e., we are interested in the distribution in $R$ of points that on scale $R$ are both upcrossing $\epsilon_c$ and are peaks.  This comoving density is given by  
\begin{align}
  \frac{{\rm d}n}{{\rm d}R}
  =  \frac{R\sigma_{23}}{7\sigma_{02}} \,\frac{p(\omega_c)}{V_*}
  \int_0^\infty \!\! &{\rm d}x\,f(x) 
  \int_{0}^\infty \!\! {\rm d}\tilde y\,\tilde y
  \,p(\tilde y,x|\omega_c)\,,
 \label{eq:expks}
\end{align}
with $\tilde y\equiv y-y_c$.  This expression combines equations \eqref{eq:dnBBKS} and \eqref{eq:ES}.

The conditional probability
 $p(\tilde y,x|\omega_c)$
can be factorized using Bayes' rule as
 $p(\tilde y|x,\omega_c)\,p(x|\omega_c)$,
where $p(x|\omega_c)$ equals the probability of $x - \gamma_{x\omega}\omega$ with $\omega=\omega_c$ (c.f. equation~\ref{eq:px|ec}), and $p(\tilde y,x|\omega_c)$ is  the conditional distribution of $y - \gamma_{y\omega}\omega$  given $x - \gamma_{x\omega}\omega$, up to a shift by $y_c$.  That is,
\begin{equation}
  p(\tilde y,x|\omega_c) = p(x|\omega_c)\,
  \frac{\exp[-(\tilde y-\mu)^2/2\Sigma^2]}{\sqrt{2\pi}\Sigma}
\label{eq:condomega}  
\end{equation}
where
\begin{align}
  b&\equiv\frac{\langle(y- \gamma_{y\omega}\omega)(x - \gamma_{x\omega}\omega)\rangle}{\langle(x - \gamma_{x\omega}\omega)^2\rangle}
  = \frac{\gamma_{xy}-\gamma_{x\omega}\gamma_{y \omega}}{1 - \gamma_{x\omega}^2}\,, \\
  \mu &\equiv\langle y|\omega_c,x\rangle -y_c
  = \gamma_{y \omega}\omega_* + b (x - \gamma_{x\omega}\omega_c)\,,
\label{eq:mu}  \\
  \Sigma^2
  &\equiv{\rm Var}(y|\omega_c,x)
  = 1-\gamma_{y \omega}^2 -b^2(1 - \gamma_{x\omega}^2) \notag\\
  &= \frac{1 - \gamma_{x\omega}^2 - \gamma_{y\omega}^2
  - \gamma_{xy}^2 + 2\gamma_{x\omega}\gamma_{y\omega}\gamma_{xy}
   }{1 - \gamma_{x\omega}^2} \,.
\label{eq:Sigma}
\end{align}
and
\begin{equation}
  \gamma_{xy}(R)\equiv\langle xy\rangle = 
  \int \frac{{\rm d}k}{k}\,\frac{k^3 P(k)}{2\pi^2}\,
  \frac{k^4\,W_3(kR)W_2(kR)}{\sigma_{23}(R)\sigma_{22}(R)}\,.
\label{eq:gammaxy}
\end{equation}

Integrating over $\tilde y$, and using the fact that ${\rm d}n/{\rm d}\ln M = (R/3)\,{\rm d}n/{\rm d}R$ because $M\propto R^3$, the excursion set peaks expression for the mass function is
\begin{align}
  \frac{{\rm d}n}{{\rm d}\ln M} =
  &\ n_*\,\frac{{\rm e}^{-\omega_c^2/2}}{\sqrt{2\pi}}
  \int_0^\infty \!\! {\rm d}x \,f(x)
  \left[\gamma_{y \omega}\omega_*+b(x-\gamma_{x\omega}\omega_c)\right] \notag \\
  &\quad\quad \times 
   p(x|\omega_c)\,
   F\bigg(\frac{\gamma_{y \omega}\omega_*+b(x-\gamma_{x\omega}\omega_c)}{\Sigma}\bigg), 
 \label{eq:dndlnm}
\end{align}
where 
\begin{equation}
  n_*\equiv \frac{R^2\sigma_{23}}{21\,\sigma_{02}\,V_*} 
  = \frac{R^2\sigma_{23}}{21\,\sigma_{02}}
  \frac{(R\sigma_{22}/\sigma_{12})^3}{(6\pi R^2)^{3/2}}
 \label{eq:n*}
\end{equation}
and the function $F$ is defined by equation \eqref{eq:F}.

In the high peaks limit, $p(x|\omega_c)$ becomes sharply peaked around its mean $\gamma_{x\omega}\omega_c$ and, since $f(x)\sim x^3$ for $x\gg1$, the integral over $x$ tends to $(\gamma_{x\omega}\omega_c)^3\gamma_{y \omega}\,\omega_*$. Moreover,
\begin{equation}
  \gamma_{y \omega}\omega_* = (\mathrm{d}\omega_c/\mathrm{d}\ln R^3)/(n_* V_*),
\end{equation}
as one can see by combining equations~(\ref{eq:w*}) and~(\ref{eq:gammayom}) for the quantities on the left hand side, and using $n_*V_*$ from equation~(\ref{eq:n*}).
As a result, 
\begin{align}
  \frac{{\rm d}n}{{\rm d}\ln M} &\to 
   \frac{(\gamma_{x\omega}\omega_c)^3}{V_*} 
  \frac{{\rm e}^{-\omega_c^2/2}}{\sqrt{2\pi}}
   \frac{{\rm d}\omega_c}{{\rm d}\ln R^3}\nonumber\\
   &= \frac{(R\sigma_{12}/\sigma_{02})^3}{(6\pi R^2)^{3/2}} 
  \frac{\omega_c^3\,{\rm e}^{-\omega_c^2/2}}{\sqrt{2\pi}}
   \frac{{\rm d}\omega_c}{{\rm d}\ln R^3}.
\label{eq:hipks}
\end{align}
The last expression shows that although the full calculation (equation~\ref{eq:dndlnm}) depends on $\sigma_{22}$, the high peaks limit does not.

Since ${\rm d}\ln M/{\rm d}\ln R^3=1$, this expression has the same form as the high peaks limit of equation~(\ref{eq:dnBBKS}) for ${\rm d}n/{\rm d}\omega_c$.  However, there $\omega_c$ denotes peak height on a fixed scale whereas here, $\omega_c=\epsilon_c/\sigma_{02}(R)$ is a proxy for the variable scale $R$ (and hence, for $M$).

Formally, peaks and excursion set peaks in the enclosed matter overdensity are still described by equation \eqref{eq:dnBBKS} and \eqref{eq:expks} respectively, replacing $\epsilon_R$ with $\delta_R$. That is, $\omega_c\to \nu_c$ and all $\sigma_{j2}\to\sigma_{j1}$.  We say `formally' because in practice, for many $P(k)$ of current interest, the integral defining $\sigma_{21}$ does not converge (whereas that for $\sigma_{22}$ does), so the calculation is inconsistent.  (E.g., the resulting expressions are ill defined since $V_\star= 0$.) For practical calculations, one has to regularize this divergence somehow, for instance by replacing the top-hat filter $W_1$ with a more convergent one (e.g. Gaussian). This tweak is not needed for our energy based approach.  (In this respect, the high peaks limit is less problematic than the full calculation, since the most divergent quantity is not required. Physically, this happens because high stationary points are likely all peaks, so the constraint on the sign of the Hessian is irrelevant.)

\subsection{A more formal derivation}
\label{sec:formal}
We give here a more elegant derivation of the results of the previous subsections, using only the transformation rule of Dirac's delta functions and treating excursion sets and peaks on the same footing.
The number density of the stationary points of a single realisation of the field $\epsilon_R(\bm{x})$ is 
\begin{equation}
  \sum_{\mathrm{sp}}\delta_{\rm D}^{(3)}(\bm{x}-\bm{x}_{\mathrm{sp}}) = |\det(\nabla\bm{\eta})|\,\delta_{\rm D}^{(3)}(\bm{\eta})\,,
\label{eq:pkpp}
\end{equation}
where $\bm{x}_{\mathrm{sp}}$ denotes each solution of the constraint equation $\eta_i(\bm{x})=0$, and $\det(\nabla\bm{\eta})=(-\sigma_{22}/\sigma_{12})^3\det(\zeta)$ is the Jacobian of the transformation from $\bm{\eta}$ to $\bm{x}$.
Similarly, the number density in $[0,+\infty)$ of the solutions $R_c$ (`crossings') of the constraint equation $\epsilon_R(\bm{x})=\epsilon_c$ is
\begin{equation}
  \sum_{\mathrm{c}}\delta_{\rm D}(R-R_c) = |\mathrm{d}(\epsilon_R-\epsilon_c)/\mathrm{d}R|\,\delta_{\rm D}(\epsilon_R-\epsilon_c)\,,
\label{eq:ESpp}
\end{equation}
where $|\mathrm{d}(\epsilon_R-\epsilon_c)/\mathrm{d}R|$ transforms the number density in $\epsilon_R$ into one in $R$.

To make equation \eqref{eq:pkpp} a density of peaks, one must count only the solutions where the eigenvalues $\zeta_i$ of $\zeta_{ij}$ are all positive. To make it a differential density of peaks of fixed height $\omega_c$, one must further multiply by $\delta_{\rm D}(\omega_R-\omega_c)$. Hence, the differential mean density of peaks of height $\epsilon_c$ is 
\begin{equation}
  \frac{{\rm d}n_{\mathrm{pk}}}{{\rm d}\omega_c} = 
  \bigg(\frac{\sigma_{22}}{\sigma_{12}}\bigg)^3
  \Bigl\langle\det(\zeta)\delta_{\rm D}^{(3)}(\bm{\eta})\delta_{\rm D}(\omega_R-\omega_c)
  \prod_i\vartheta(\zeta_i)\Bigr\rangle \,,
\end{equation}
where the angle brackets denote an average over the joint distribution of $\omega,{\bm \eta},{\bm \zeta}$ on scale $R$.  This average gives equation \eqref{eq:dnBBKS} for a fixed smoothing scale $R$. Integrating this over $\epsilon_c$ gets rid of $\delta_{\rm D}(\epsilon_R-\epsilon_c)$, since integration and mean commute.

Similarly, to make equation \eqref{eq:ESpp} a density of upcrossing points, the derivative $\mathrm{d}(\omega_R-\omega_c)/\mathrm{d}R$ must be negative.  The mean distribution in $R$ of such points is thus
\begin{equation}
  f_{\mathrm{up}}(R) = 
  \Bigl\langle|\omega'-\omega_c'|\,
  \delta_{\rm D}(\omega_R-\omega_c)\,
  \vartheta(\omega_R'-\omega_c')\Bigr\rangle,
\end{equation}
where $'=\mathrm{d}/\mathrm{d}R$, and the angle brackets denote an average over the joint distribution of $\omega_R$ and $\omega_R'$.  Note that $\langle\omega\omega'\rangle=0$ (since $\omega$ is normalized) and direct calculation shows that $\omega'\propto y-\gamma_{y\omega}\omega$ is the relevant quantity for $p(y|\omega_c)$.  Thus, the result is equation \eqref{eq:ES} up to a total derivative $\mathrm{d}\sigma_{02}/\mathrm{d}R$ to change variables from $R$ to $\sigma_{02}(R)$. 

When combining peaks and excursion sets, the problem becomes that of finding the density of points in the 4-dimensional space $\{R,\bm{x}\}$ that solve $\{\omega,\bm{\eta}\}=\{\omega_c,\bm{0}\}$ (plus the peak and upcrossing constraints). That is,
\begin{equation}
  \frac{{\rm d}n}{{\rm d}R} \equiv 
  \Bigl\langle |J|\, 
  \delta_{\rm D}^{(3)}(\bm{\eta})\,
  \delta_{\rm D}(\omega_R-\omega_c)
  \prod_i\vartheta(\zeta_i)\,
  \vartheta(\omega_R'-\omega_c')\Bigr\rangle,
  \label{eq:espJ}
\end{equation}
where $J = \det[\partial\{\omega_R-\omega_c,\bm{\eta}\}/\partial\{R,\mathbf{x}\}]$ is the 4-dimensional Jacobian determinant.
Since $\nabla(\omega_R-\omega_c)=0$ at peaks, $J$ factorizes into $|J|=(\sigma_{22}/\sigma_{12})^3|\det(\zeta)||\mathrm{d}(\omega_R-\omega_c)/\mathrm{d}R|$. Equation~\eqref{eq:espJ} is the same as equation~\eqref{eq:expks}.

\subsection{Peak velocities and velocity bias}
We have emphasized the fact that our energy-peaks excursion set approach is a simple but well-motivated modification of the density-based approach.  As \cite{bbks86} noted, the peaks constraint modifies the distribution of velocities.  This is because, although velocities do not correlate with (even space derivatives of) the density, they do correlate with odd derivatives.

For Gaussian statistics, the constrained mean is linear in the constrained variables.  Since the gradient vanishes at a peak, the mean peak velocity is zero (in each direction).  However, the dispersion around this mean is modified by the constraint on the gradient.  Since the Fourier transform of the velocity is given by ${\bm v} = i{\bm k}\,\delta({\bm k})/k^2\,W_1(kR)$ and the gradient of $\epsilon_R$ is ${\bm \nabla}\epsilon = i{\bm k}\,\delta({\bm k})\,W_2(kR)$, their cross-correlation is 
\begin{equation}
  \langle {\bm v}\cdot{\bm \nabla}\epsilon\rangle =
  \int \frac{{\rm d}k\,k^2}{2\pi^2}\,P(k)\,W_1(kR)W_2(kR) = -\langle\delta\epsilon\rangle
\end{equation}
and the peak velocity dispersion is 
\begin{equation}
  \sigma^2_{{\rm vpk}} = \sigma^2_{-11}\,\left(1 - \frac{\langle {\bm v}\cdot{\bm \nabla}\epsilon\rangle^2}{\sigma^2_{-11}\,\sigma^2_{12}}\right).
 \label{eq:rmsvpk}
\end{equation}
Thus,
\begin{equation}
  f(v)\,{\rm d}v = \frac{\exp(-3v^2/2\sigma^2_{\rm vpk})}{(2\pi\sigma^2_{\rm vpk}/3)^{3/2}}\, 4\pi v^2\,{\rm d}v \,.
\end{equation}
Since $\sigma^2_{\rm vpk} \ne \sigma^2_{-11}$ this manifests as `velocity bias' in Fourier space \citep{ds10}.  `Velocity bias' is unfortunate nomenclature because it suggests that the velocity of a peak identified on scale $R$ differs from the mean motion of the dark matter within $R$, when in fact this is the definition of the speed of the peak patch.  The 'bias' is really with respect to the statistics of dark matter speeds and arises from two separate effects:  the dependence on smoothing scale $R$, and the peak constraint, which gives rise to the terms which correct $\sigma_{-11}$.

\subsection{Scale dependence and scatter in $\delta_R$}\label{sec:scat}
Because of equation \eqref{eq:yepsdel}, $y$ and $\delta$ are deterministically (and linearly) related at fixed $\epsilon$. Thus, a scatter in $y$ automatically produces a scatter in $\delta$. In particular, the integral over $\tilde y$ in equation \eqref{eq:expks} shows explicitly that peaks which are upcrossing on the same $R$ can have a range of $\delta$, weighted by $p(\tilde y|x,\epsilon_c)$.
That is to say, a model in which haloes of the same mass all have the same $\epsilon_c$ generically predicts scatter in the enclosed overdensities of the protohalo patches, with {\em no} additional work (i.e., equation \ref{eq:expks} is no more complicated than the excursion set peaks model for $\delta$, in which $\delta_c$ is deterministic rather than stochastic).  

\begin{figure}
  \includegraphics[width = \columnwidth]{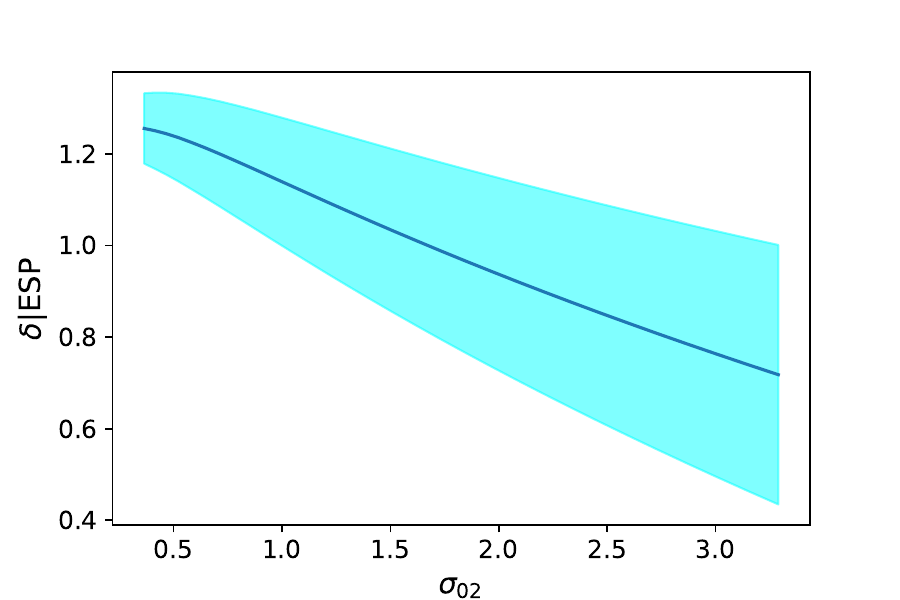}
  \caption{\label{fig:conddelta} Conditional mean (equation~\ref{eq:dpk}) plus and minus twice the standard deviation (square root of equation~\ref{eq:vpk}) of $\delta_R$ for excursion set peaks in energy having $\epsilon_c=1.686$ for all masses (parametrized by $\sigma_{02}$ of equation~\ref{eq:sigman}).}
\end{figure}

The moments of the conditional distribution of $\delta$ given the ESP constraint (peaks plus upcrossing) on scale $R$ can be obtained inserting powers of
\begin{equation}
  \delta = \epsilon_c + \frac{{\rm d}\epsilon_c}{{\rm d}\ln R^5}
  - \frac{R^2\sigma_{23}}{35}\tilde y
\label{eq:delta}
\end{equation}
in equation \eqref{eq:expks}, and then dividing by $\mathrm{d}n/\mathrm{d} R$.
For instance, the mean value and variance of $\delta$ at fixed $R$ are
\begin{align}
  &\langle\delta|{\rm ESP}\rangle = \epsilon_c
  + \frac{{\rm d}\epsilon_c}{{\rm d}\ln R^5}
  - \frac{R^2\sigma_{23}}{35} 
  \, \frac{G_2(\epsilon_c,y_c)}{G_1(\epsilon_c,y_c)}\,,
  \label{eq:dpk} \\
  & \mathrm{Var}(\delta|\mathrm{ESP}) =
  \bigg(\frac{R^2\sigma_{23}}{35}\bigg)^2 
    \, \bigg[ \frac{G_3(\epsilon_c,y_c)}{G_1(\epsilon_c,y_c)}
    -\frac{G_2^2(\epsilon_c,y_c)}{G_1^2(\epsilon_c,y_c)}\bigg]\,,
   \label{eq:vpk}
\end{align}
where 
\begin{equation}
  G_n(\epsilon_c,y_c) \equiv 
  \int_0^\infty \!\! {\rm d}x\,f(x) \,p(x|\epsilon_c)
  \int_{0}^\infty \!\!{\rm d}\tilde y\,\tilde y^n\,
  p(\tilde y|x,\epsilon_c)
\label{eq:Gn}
\end{equation}
and $p(\tilde y|x,\epsilon_c)$ is given by equation \eqref{eq:condomega}.

For any $n$, the integral over $\tilde y$ in equation \eqref{eq:Gn} equals 
\begin{align}
  \frac{(\sqrt{2} \Sigma)^n}{\sqrt{\pi}}
  \bigg[&
  \frac{\mu}{\sqrt{2}\Sigma} \Gamma\bigg(\!1 + \frac{n}{2}\bigg)
  {}_1F_1\bigg(\frac{1 - n}{2}, \frac{3}{2},-\frac{\mu^2}{2 \Sigma^2}\bigg)
  \notag\\
  &+ \frac{1}{2} \,\Gamma\bigg(\frac{1 + n}{2}\bigg) 
  {}_1F_1\bigg(-\frac{n}{2}, \frac{1}{2},
  -\frac{\mu^2}{2 \Sigma^2}\bigg)\bigg]
\end{align}
where ${}_1F_1(a,b,z)$ is the confluent hypergeometric function, and $\mu$ and $\Sigma$ are the mean and variance of $p(\tilde y|x,\epsilon_c)$, given by equations~\eqref{eq:mu} and~\eqref{eq:Sigma} respectively. For $n=1$ this returns $\mu\, F(\mu/\Sigma)$, with $F$ defined as in equation \eqref{eq:F}; for $n=2$ it gives
\begin{equation}
  \frac{\mu^2+\Sigma^2}{2}
  \bigg[1+\mathrm{erf}\bigg(\frac{\mu}{\sqrt{2}\Sigma}\bigg)\bigg]
  + \mu^2\frac{\mathrm{e}^{-(\mu/\Sigma)^2/2}}{\sqrt{2\pi}(\mu/\Sigma)}\,;
\end{equation}
and for $n=3$
\begin{equation}
  \frac{\mu^3+3\mu\Sigma^2}{2}
  \bigg[1+\mathrm{erf}\bigg(\frac{\mu}{\sqrt{2}\Sigma}\bigg)\bigg]
  + (\mu^3+2\mu\Sigma^2)\frac{\mathrm{e}^{-(\mu/\Sigma)^2/2}}{\sqrt{2\pi}(\mu/\Sigma)} \,.
\end{equation}

The high peak limit of these expressions is recovered by setting $\mu\simeq\gamma_{y\omega}\omega_*\gg\Sigma$, in which case one gets that $\langle\delta|\mathrm{ESP}\rangle\to\epsilon_c(1+\mathrm{d}\ln\sigma_{02}/\mathrm{d}\ln R^5)<\epsilon_c$ and $ \mathrm{Var}(\delta|\mathrm{ESP}) \to (R^2\sigma_{23}\Sigma/35)^2$. 

We show in the Appendix that $R^2\sigma_{23} \propto \sigma_{02}$, with a correction factor that is independent of $R$ for power law power spectra.  In addition, $\sigma_{02}\propto \sigma_{01}$ for such spectra, so we generally expect rms$(\delta|\mathrm{ESP})$ to be approximately $\propto \sigma_{01}$, which is in agreement with the scatter around the mean overdensity of protohaloes identified in simulations \citep{dwbs08,robEtal08,Elia2011,dts13}.
Figure~\ref{fig:conddelta} shows a plot of $\langle\delta|\mathrm{ESP}\rangle$ with the rms spread around this mean, for $\epsilon_c = 1.686$.  Notice that the mean is smaller than $\epsilon_c$, a consequence of the upcrossing constraint that $y_R\propto (\epsilon_R-\delta_R)\ge 0$.  Moreover, the mean is a decreasing function of $\sigma_{02}$, and hence of $\sigma_{01}$.  In contrast, the mean overdensity within protohalo patches identified in simulations increases approximately as $1.686 + 0.4\sigma_{01}$ \citep{smt01,scs13}.  Therefore, we might naively expect $\epsilon_c$ to increase even more strongly with $\sigma_{02}$.  We test this prediction below.

\section{Comparison with simulations}
\label{sec:sims}
This section describes a few simple tests of our formalism using 5378 haloes identified in the $z=0$ outputs of the Flora simulation, the largest box in the SBARBINE suite \citep{despali16}. Our halo set is composed as follows: all the 1378 halos more massive than $10^{15}h^{-1}M_\odot$, 2000 randomly chosen haloes with mass between $10^{14}$ and $ 10^{15}h^{-1}M_\odot$, and 2000 randomly chosen haloes with mass between $4\times 10^{13}$ and $10^{14}h^{-1}M_\odot$.

The simulation followed the gravitational evolution of $1024^3$ dark matter particles each of mass $6.35\times10^{11} h^{-1}M_\odot$ in a periodic cube of side $L_{\mathrm{box}}=2h^{-1}$Gpc with a Planck13 background cosmology: $\Omega_m = 0.307$, $\Omega_\Lambda = 0.693$, $\sigma_8 = 0.829$ and $h = 0.677$.

\begin{figure}
  \includegraphics[width = \columnwidth]{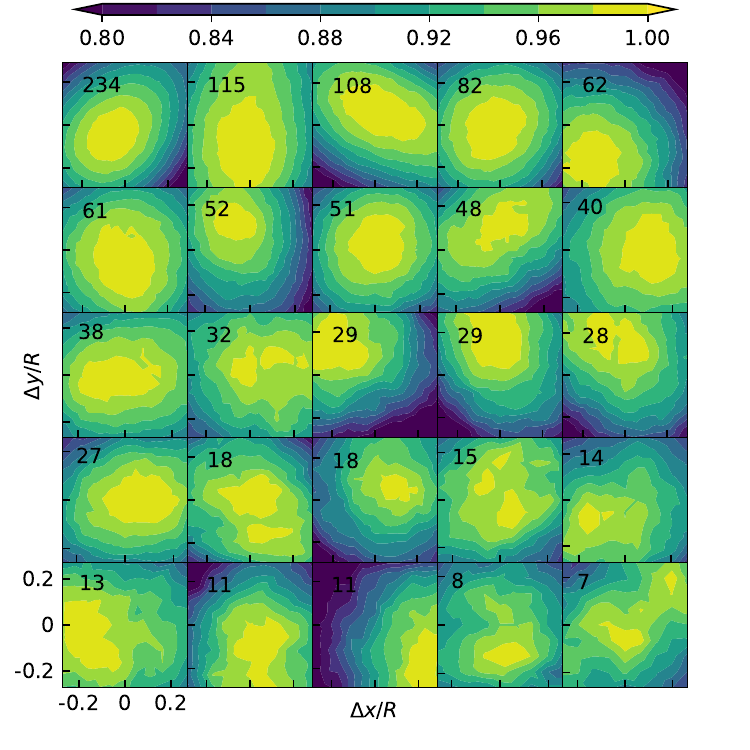}
  \caption{Maps of the $\epsilon_R$ field, each normalised to its maximum value, in the central regions $(|\bm{x}-\rcm|\le 0.2R)$ of protohaloes:  brighter shade indicates larger $\epsilon_R$.  The integer in the top left corner of each panel shows the mass within $R$ in units of $10^{13}h^{-1}M_\odot$.}
  \label{fig:eps}
\end{figure}
\begin{figure}
  \includegraphics[width = \columnwidth]{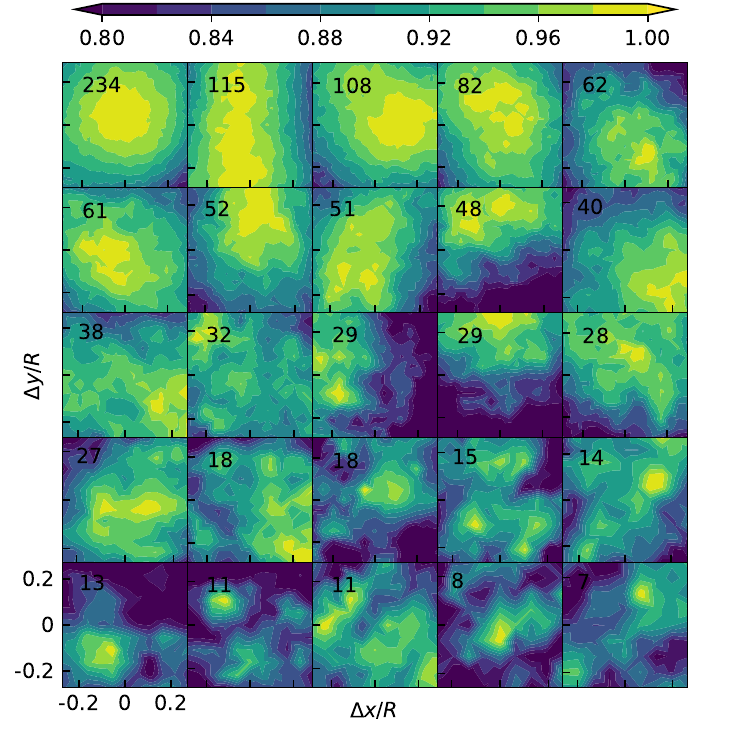}
  \caption{Same as Figure~\ref{fig:eps} but for $\delta_R$. Compared to $\epsilon_R$, the $\delta_R$ field has less distinctive peaks which are more offset from the protohalo center, and more substructure which degrades the peak-protosphere correspondence.}
  \label{fig:del}
\end{figure}

\begin{figure*}
  \centering
  \includegraphics[width= \columnwidth]{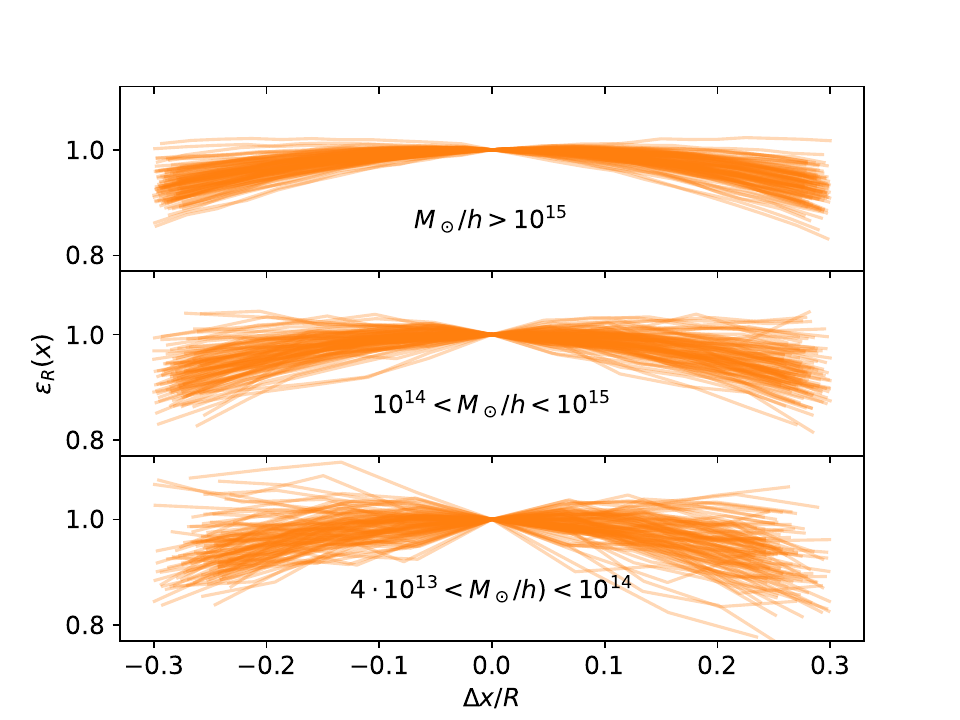}
  \includegraphics[width= \columnwidth]{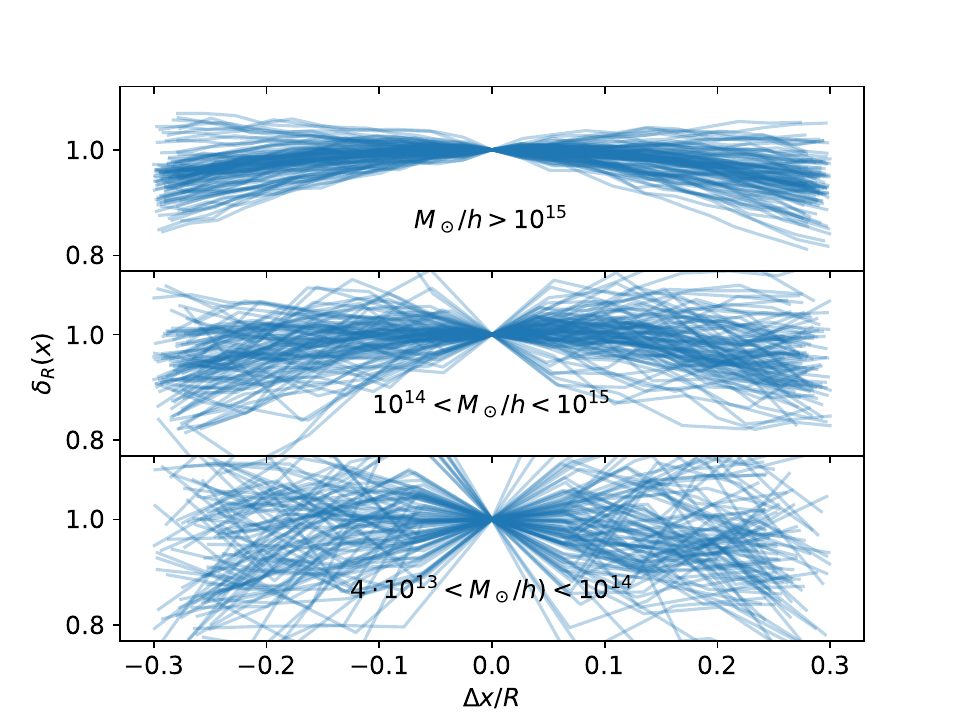}
  \caption{Profiles of $\epsilon_R$ (left) and $\delta_R$ (right) in the $x$ direction, normalized to the central value, for 100 randomly chosen haloes in three different mass bins, as a function of the displacement of the protosphere from the protohalo's center of mass.}
  \label{fig:xprofiles}
\end{figure*}

Haloes were identified using a Spherical Overdensity halo finder with threshold of $319\times$ the background density.  We use `protohalo' to refer to the patch defined by a halo's particles in the initial conditions.  This patch is not spherical, whereas our formalism is based on spherically symmetric smoothing filters. To measure spherically averaged quantities in the initial conditions, for each protohalo having e.g. $N_{\rm halo}$ particles we identify a `protosphere': this is the set of all particles closer than $R=(3N_{\rm halo}/4\pi)^{1/3}(L_{\rm box}/1024)$ to the protohalo's center of mass.

\subsection{Estimators}
In practice, we compute averages by explicitly summing over real space positions, rather than using Fourier methods.  This means that we estimate
\begin{equation}
  \hat\epsilon_R = -3\, \frac{\sum_j [(\bm{r}-\rcm)\cdot(\bm{v}-\bm{v}_{\mathrm{cm}})/fDH]_j}{\sum_j [(\bm{r}-\rcm)\cdot(\bm{r}-\rcm)]_j} ,
 \label{eq:este}
\end{equation}
where the sum is over the  particles in the protohalo or in the protosphere (compare equation~\ref{eq:eps}, and recall that $\nabla\phi = -v/fH$ initially).  Note that the denominator is close to $(3/5)\,N_{\mathrm{halo}}R^2$.  Similarly, we estimate
\begin{equation}
  \hat\delta_R = -3\frac{\sum_{j\in\mathrm{sh}} [(\bm{r}-\rcm)\cdot(\bm{v}-\bm{v}_{\mathrm{cm}})/fDH]_j}{\sum_{j\in\mathrm{sh}} [(\bm{r}-\rcm)\cdot(\bm{r}-\rcm)]_j}
 \label{eq:estd}
\end{equation}
where now $j$ runs only over the particles in a thin spherical shell of radius $R$ (e.g. $0.95R<|\bm{r}-\rcm|<1.05R$).  (We only retain haloes with more than 10 particles in this shell.) We have checked that this measurement agrees with $\hat\epsilon_R + ({\rm d}\hat\epsilon_R/{\rm d}\ln R)/5$ (see equations~\ref{eq:slopeEps} and~\ref{eq:yepsdel}) where the derivative is estimated by first evaluating equation~(\ref{eq:este}) for many narrowly spaced $R$ values (while keeping $\rcm$ and $\bm{v}_{\mathrm{cm}}$ fixed to their values for the full sphere).

\subsection{Peaks in the initial $\epsilon_R$ and $\delta_R$ distributions}
Our first test is to check if $\epsilon_R$ is (close to) a local maximum. For this, we generate a fine cartesian grid with spacing $0.02R$, out to $\pm 0.2R$ from the center of mass position. We measure $\epsilon_R$ at each grid point using equation~(\ref{eq:este}), where $j$ labels all particles within $R$ from the grid point.  Figure~\ref{fig:eps} shows the $\epsilon_R$ field in the central regions (i.e. within $0.2R$ in the $x-y$ plane at $z=0$) of 25 randomly selected protohaloes, ordered by mass.  The integers in the top left panels show the protohalo's mass in units of $10^{13}h^{-1}M_\odot$.  The color scale is normalized to the maximum value of $\epsilon$ in each panel.  It is clear that the most massive objects (top) are noticeably better centered and smoother than the lower mass ones (bottom).

Maps of $\delta_R$ for these same regions are shown in Figure~\ref{fig:del}.  While it is again true that the most massive objects are better centered and smoother, these maps are clearly much less smooth, and not nearly as well centered as the corresponding maps of $\epsilon_R$.  The difference is particularly striking at the lower masses shown in the bottom two or three rows.

To emphasize this point, Figure~\ref{fig:xprofiles} shows the ratio $\epsilon_R(\Delta x)/\epsilon_R(0)$ (left) and $\delta_R(\Delta x)/\delta_R(0)$ (right) along the $x$-axis, where $\Delta x$ is the offset from the protohalo's center of mass, for 100 randomly selected protohaloes.  Note that the $\epsilon_R$ profiles are more likely to curve downwards.  They also tend to be smoother; the $\delta_R$ profiles can have several peaks within a small distance of the center of mass position, consistent with the fact that the variance of $\nabla^2\delta_R$ diverges.  This illustrates why $\epsilon_R$ is better suited for identifying protohaloes.  

\begin{figure}
  \includegraphics[width = \columnwidth]{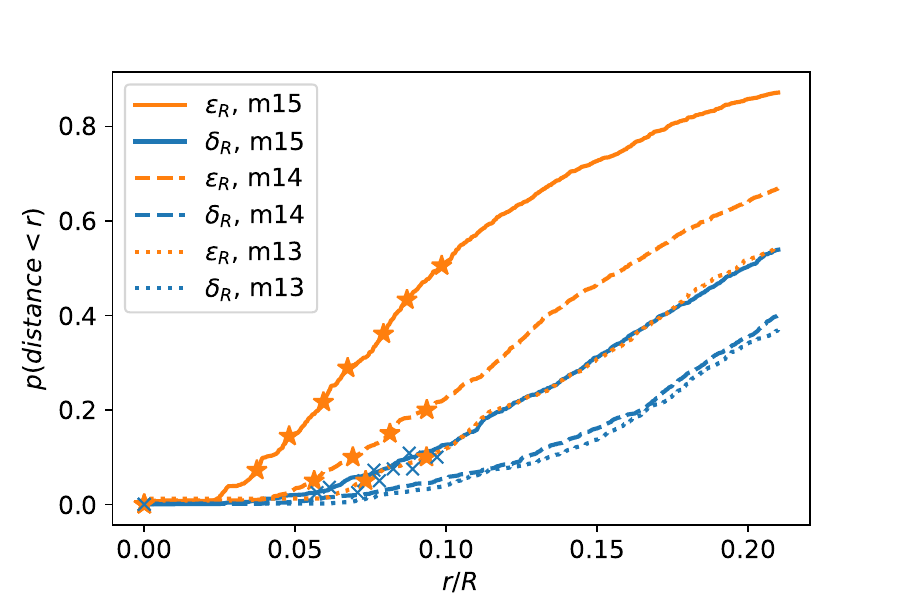}
  \caption{Fraction of protospheres for which the maximum value of $\epsilon_R$ (solid orange) or $\delta_R$ (dashed blue) lies closer than $r$ to the center of mass, shown as a function of $r/R$. Curves are calculated using all protospheres centered at less than $0.3R$ from the protohalo's center of mass. Symbols show the same for a distance of less than $0.2R$. They indicate good convergence for $\epsilon_R$ but not for $\delta_R$}
  \label{fig:cumulative}
\end{figure}

Finally, Figure~\ref{fig:cumulative} shows an estimate of the correspondence between the local maximum in $\epsilon_R$ or $\delta_R$ and the protohalo center.  For each protohalo, we first smooth $\epsilon$ with a filter of scale $R$, at a number of positions within $0.3R$ of the protohalo center.  If our model is correct, then the position of the largest of these smoothed $\epsilon_R$ values should coincide with the protohalo center.  The orange curves in Figure~\ref{fig:cumulative} show the cumulative distribution of the distance (in units of $R$) between the position of the largest value within $0.3R$ and the protohalo center, for the halos in our three broad mass bins.  (We only show scales $r\le 0.25R$ because all the curves must reach unity at $0.3R$, by definition.)  There is a clear trend with mass -- massive halos are better centered -- so this quantifies the obvious mass dependence in Figure~\ref{fig:eps}.  Note that $r/R=0.215$ corresponds to the innermost 1\% of the volume, so the fact that more than 80\% of the massive objects have their maximum $\epsilon_R$ in the innermost 1\% of the volume is remarkable.  For lower masses, this fraction is approximately 50\%.  As a converence check, we have repeated the exercise, but now restricting our search for a local maximum to $0.2R$.  In this case, all the curves must reach unity at $0.2R$, so the symbols only show the distributions out to $0.1R$.  They lie on the smooth curves, indicating convergence.

  The blue curves in Figure~\ref{fig:cumulative} show a similar analysis of $\delta_R$.  Comparison with the orange curves shows that, for all three mass bins, the maximum in $\delta_R$ tends to lie substantially further from the center than it does for $\epsilon_R$.  The blue crosses show the result of restricting our search for a local maximum to $0.2R$ (rather than $0.3R$), for the halos in the middle mass bin.  It is clearly offset from the dashed blue curve, indicating that the procedure has not converged.  Had we repeated the analysis using maxima within $0.4R$ instead, the curves would all have smaller values of $p(<r/R)$, at least for $r/R<0.1$ or 0.2.  These lower values, and the lack of convergence, are yet another way of illustrating why $\epsilon_R$ is better suited for identifying protohaloes than is $\delta_R$.

\subsection{Collapse thresholds, slope and stochasticity}
\begin{figure}
  \includegraphics[width=\columnwidth]{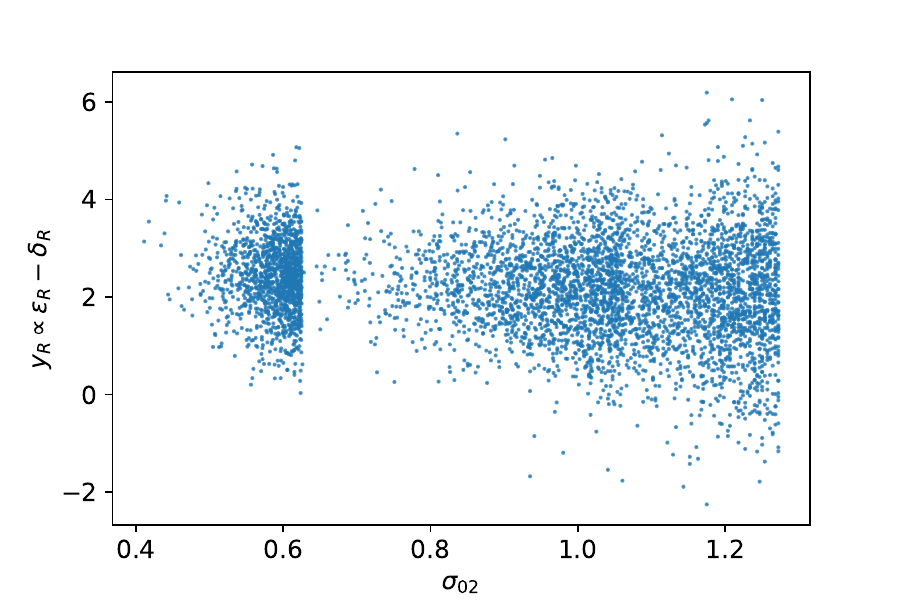}
  \caption{\label{fig:diff}Difference between spherically averaged $\epsilon_R$ and $\delta_R$, scaled by $R^2\sigma_{23}/35$ so as to match the excursion set variable $y_R$ of equation~\eqref{eq:yepsdel}, plotted as a function of $\sigma_{02}$.}
\end{figure}

Our next test is to check if the excursion set quantity $y_R\propto \epsilon_R-\delta_R$ (the slope of the  trajectory, c.f. equation~\ref{eq:yepsdel}) is strictly positive.  Figure~\ref{fig:diff} shows that it is:  $y_c\approx 2\pm 0.7$ for all protospheres.

\begin{figure}
  \centering
  \includegraphics[width=\columnwidth]{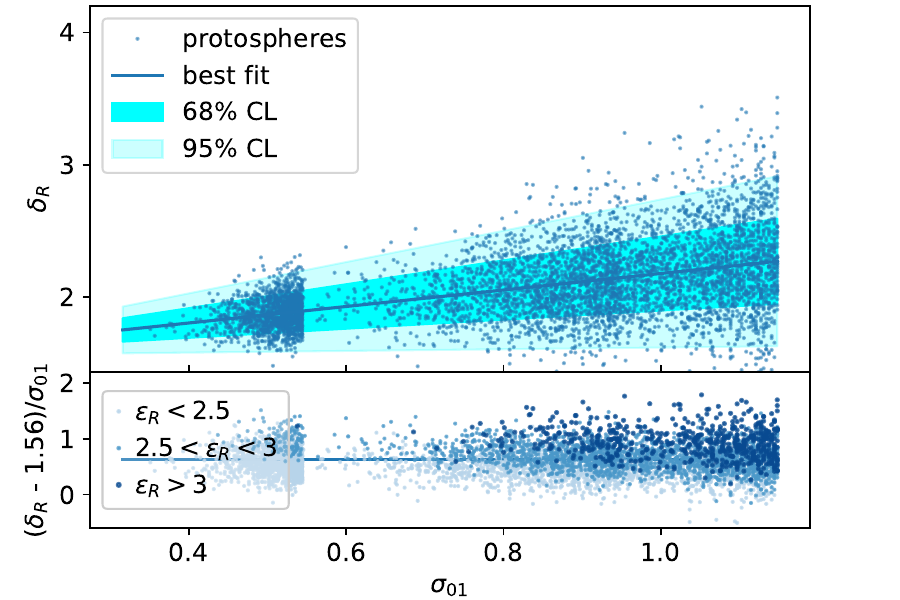}
  \includegraphics[width=\columnwidth]{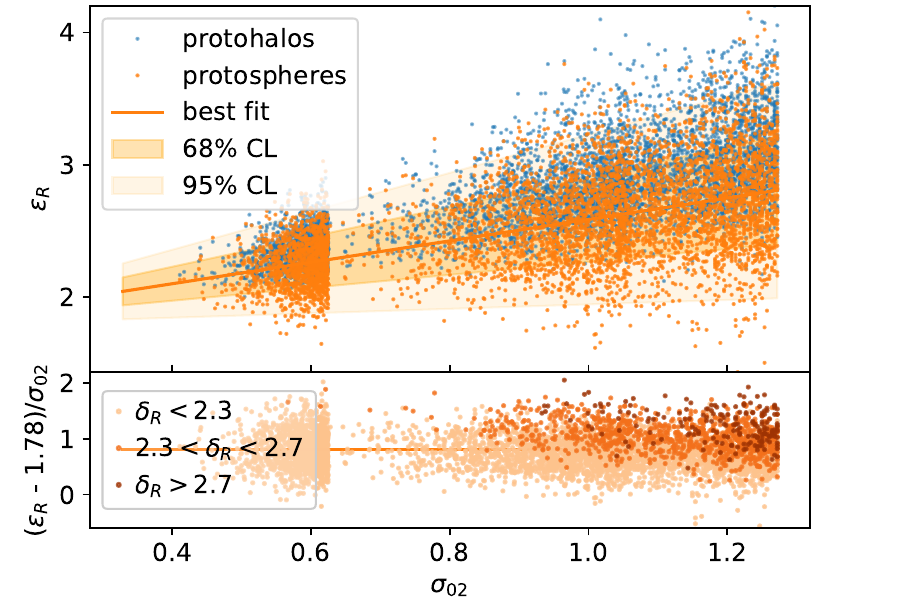}
  \caption{\label{fig:epsc} Mean matter (top) and energy (bottom) overdensity in protospheres (spheres of Lagrangian radius $R$ centered on the protohalo's center of mass) and in protohaloes (for energy only), shown as a function of $\sigma_{01}$ and $\sigma_{02}$ respectively. Solid lines and shaded regions show best fits (equations~\ref{eq:fitd} and~\ref{eq:fite}) and the 68\% and 95\% prediction bands. Lower panels show these results, scaled to remove the trend with $\sigma_{01}$ and $\sigma_{02}$, with darker (lighter) colors indicating larger (smaller) values of the other variable as labeled.} 
\end{figure}

Following Section~\ref{sec:scat}, yet another test is to check if $\epsilon$ is a stronger function of protohalo mass than is $\delta$.  The top panel of Figure~\ref{fig:epsc} shows $\delta_R$ in the protosphere of radius $R$ centered on the protohalo center of mass for a number of protohaloes, each labeled by its $\sigma_{01}(R)$ (recall that $M\propto R^3$ increases as $\sigma_{01}(R)$ decreases).  The solid line shows the linear regression
\begin{equation}
  \delta_R  = 1.56 + 0.63\,\sigma_{01},
\label{eq:fitd}
\end{equation}
which provides the best fit to the mean trend.  This is qualitatively consistent with, although steeper than, the 0.48 scaling reported in previous literature \citep[e.g.][]{smt01,robEtal08}.

The rms scatter around the mean scales approximately as $0.32\,\sigma_{01}$, and is also consistent with previous work \citep{robEtal08}.  The dark and light shaded regions show the mean plus and minus one or two standard deviations (i.e., the 68\% and 95\% prediction bands).
The bottom part of this panel shows the data in the format used by \cite{dts13}; this removes the scaling of the mean and rms with $\sigma_{01}$ so the mean height, 0.63, equals the slope of equation~(\ref{eq:fitd}) and the amplitude of the rms scatter around it gives the scaling of the rms with $\sigma_{01}$, 0.32. In this panel, darker colors show objects that have larger values of $\epsilon_R$: as expected, $\epsilon_R$ and $\delta_R$ are strongly correlated. 

In the lower panel of Figure~\ref{fig:epsc}, the orange symbols show a similar analysis of $\epsilon_R$ as a function of $\sigma_{02}$.  The solid line shows the linear regression 
\begin{equation}
  \epsilon_R  = 1.78 + 0.81\, \sigma_{02}.
\label{eq:fite}
\end{equation}
This is steeper than equation~\eqref{eq:fitd}, in agreement with the discussion in Section~\ref{sec:scat}.  The rms scatter around this relation increases as mass decreases (qualitatively like for $\delta_R$:  in this case it scales as $0.32\,\sigma_{02}$.  
The blue symbols in the panel show the result of estimating $\epsilon_R$ using the protohalo (rather than protosphere) particles.  They tend to lie above the orange ones, consistent with the fact that each protosphere likely contains some particles that did not make it into the halo and are therefore likely to be less bound than the full set of protohalo particles.  The bottom part of the panel shows that the slope of the $\epsilon_R-\sigma_{02}$ relation is 0.8, and the differently colored symbols show that the scatter correlates with $\delta_R$.  As a result, at fixed $\delta_R$ the scatter in $\epsilon_R$ is smaller.

\begin{figure}
  \includegraphics[width=\columnwidth]{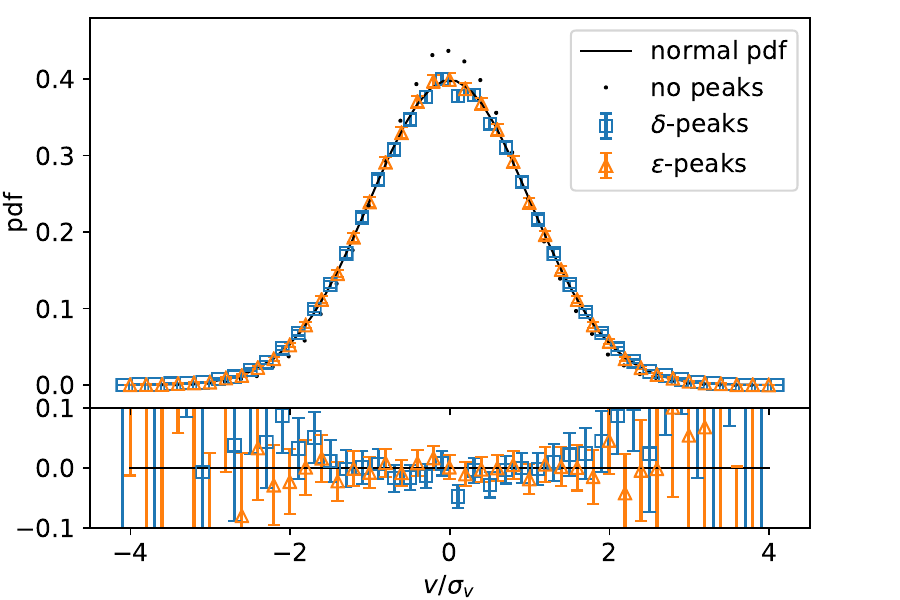}
  \caption{\label{fig:pdfv} Distribution of scaled protosphere velocities.  
    Triangles show the result of scaling all $v$ by equation~\eqref{eq:rmsvpk} (the $\epsilon_R$-peaks scaling);
    squares scale by equation~\eqref{eq:rmsvpk} after replacing all occurences of $W_2$ in it by $W_1$ (the $\delta_R$-peaks scaling); 
    dots scale by $\sigma_{-11}$ (the first term in equation~\ref{eq:rmsvpk});
    solid black line shows a Gaussian with zero mean and unit rms.}
\end{figure}

\subsection{Velocities}
Finally, we have compared the velocities of protosphere and protohalo patches with our predictions.  The rms velocity is smaller for more massive protospheres, and is always smaller than $\sigma_{-11}$.  This is in qualitative agreement with previous measurements which showed that the prediction based on $\delta_R$-peaks (set all $W_2\to W_1$ in our equation~\ref{eq:rmsvpk}) is quite accurate \citep{sd01}.  The $\delta_R$- and $\epsilon_R$-peak predictions only differ by a few percent so, to increase signal to noise, Figure~\ref{fig:pdfv} shows the distribution of $v_{\rm 1d}/\sigma_{\rm 1d}$ in protospheres, where $v_{\rm 1d}$ represents each of the three cartesian components of the velocity, and $\sigma_{\rm 1d}^2 = \sigma_{\rm vpk}^2/3$.  If the mass dependence of $\sigma_{\rm 1d}$ is correct, the resulting distribution should be Gaussian with zero mean and unit rms.

The smooth black curve shows such a unit variance Gaussian.  Orange triangles and blue squares show our measurements when $\sigma_{\rm 1d}$ is that for $\epsilon_R$- and $\delta_R$-peaks; they are very similar to the Gaussian.  (Statistical errors on the measurements are similar to the symbol sizes shown.)  The black dotted curve uses $\sigma_{\rm 1d} = \sigma_{-11}$ which includes the effect of smoothing  the velocities but not the additional terms coming from the peaks constraint: it is clearly inconsistent with the Gaussian shape.  As results for protohaloes are almost identical, we conclude that the peaks-based approach provides a good description of protohalo speeds, but the differences between $\epsilon_R$- and $\delta_R$-peaks are too small to matter.

\subsection{Halo abundances: An illustrative example}
\begin{figure}
  \includegraphics[width=\columnwidth]{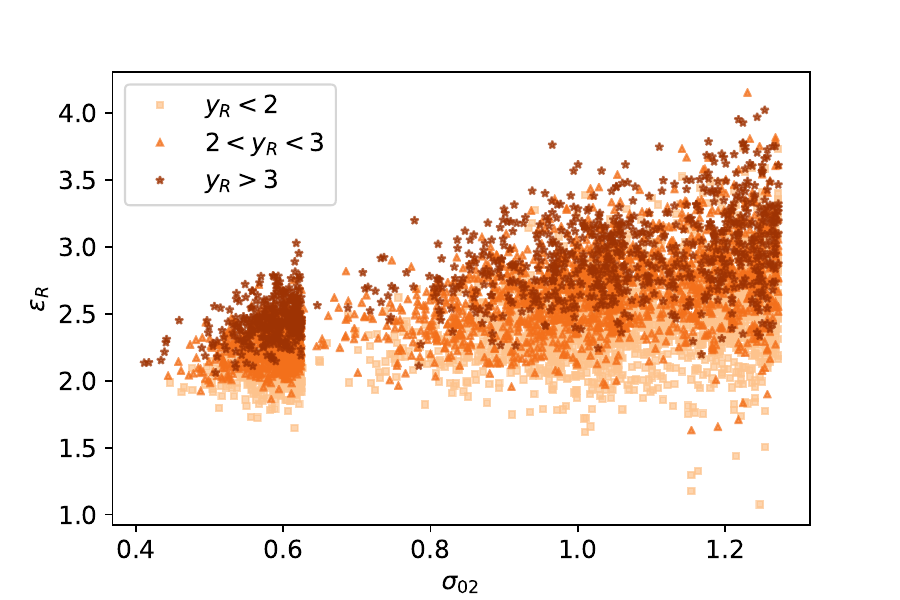}
  \caption{\label{fig:ey-sig} Dependence of $\epsilon_R-\sigma_{02}$ correlation on excursion set slope parameter $y_R\propto (\epsilon_R-\delta_R)$.}  
\end{figure}

This subsection presents a comparison of the halo mass function measured in the $z=0$ output of the simulation with that predicted by our approach.  The simplest prediction, equation~(\ref{eq:dndlnm}), requires knowledge of how $\epsilon_c$ depends on smoothing scale $R$ (and hence on mass).  We could use equation~(\ref{eq:fite}) for this, but, as Figure~\ref{fig:epsc} shows, there is substantial scatter (rms$=0.3\,\sigma_{02}$) around the mean trends with mass.  Moreover, this scatter correlates with $\delta_R$, and hence with $y_R\propto\epsilon_R-\delta_R$.  Figure~\ref{fig:ey-sig} shows this explicitly:  objects with steeper slopes tend to have larger $\epsilon_R$.  At fixed $y_R$, the scatter is reduced slightly to rms$=0.25\,\sigma_{02}$.  

To include this correlation we assume that 
\begin{equation}
  \label{eq:ecyq}
  \epsilon_c(y_R,q_R) = a_0 + a_1\,\sigma_{02}(R)\, y_R + a_2\, q_R
\end{equation}
where $q_R$ is a random variate that we assume is uncorrelated with either of $\epsilon_R,y_R,x_R$ and accounts for the scatter in $\epsilon_c$ at fixed $y_R$ shown in Figure~\ref{fig:ey-sig}.  This is motivated by previous work with $\delta_R$, in which $q_R$ is related to the amplitude of the shear field \citep{st02, scs13}.  The shear statistics are $\chi^2$ with 5 degrees of freedom.  To highlight the fact that we have not yet built shear statistics into our model, rather than using $\chi^2_5$ statistics, we assume that $q_R =\sigma_{02}(R)\,q$, where $q$ is drawn from a Lognormal with unit mean and rms chosen to match Figure~\ref{fig:ey-sig}; this makes our $q_R$ more like the parameter $\beta$ in previous ESP work with $\delta_R$ \citep[e.g.][]{psd13}.  Matching Figure~\ref{fig:ey-sig} requires $(a_0,a_1,a_2)=(1.71,0.21,0.42)$ and the rms of $q$ to be 0.25.

Then, equation~\eqref{eq:expks} -- or \eqref{eq:dndlnm} -- should be replaced by
\begin{equation}
  \frac{{\rm d}n}{{\rm d}\ln M} = \int dq\,p(q)
  \frac{{\rm d}n(M|q)}{{\rm d}\ln M},   
  \label{eq:expksq}
\end{equation}
where the mass function at fixed $q$ is
\begin{equation}
  \frac{{\rm d}n(M|q)}{{\rm d}\ln M} =
  n_* \!\int_0^\infty \!\!\!\! {\rm d}x\,f(x) \!
  \int_Y \!{\rm d}y\,(y-y_c)\,p(y,x,\omega_c(y,q)),
  \label{eq:expksq2}
\end{equation}
$\omega_c$ is $\epsilon_c$ of equation~(\ref{eq:ecyq}) divided by $\sigma_{02}$, and $Y$ is the domain where $y-y_c>0$. Note that now equation~\eqref{eq:wcyc} gives
\begin{equation}
  y_c = - \frac{7}{R^2\sigma_{23}}\bigg(a_1 y_R \frac{\mathrm{d}\sigma_{02}}{\mathrm{d}\ln R} + 
  \frac{\mathrm{d}\epsilon_c}{\mathrm{d}y_R}
  \frac{\mathrm{d}y_R}{\mathrm{d}\ln R}
  + \frac{\mathrm{d}\epsilon_c}{\mathrm{d}q_R}
  \frac{\mathrm{d}q_R}{\mathrm{d}\ln R}\bigg)
  \label{eq:ycscatter}
\end{equation}
with\footnote{These expressions assume that the derivative acts only on the explicit $R$-dependence in $y_R(\bm{x})$ and $q_R(\bm{x})$, and not on the scale dependence of the peak position $\bm{x}$. We show in Appendix \ref{app:alt} that this is equivalent to assuming that $\bm{x}$ is also a peak for $\delta_R$; otherwise, an additional term $(-7/R^2\sigma_{23})(5/\sigma_{22})\,(\nabla_i\delta\, \zeta_{ij}^{-1}\nabla_j\delta$) would appear (see equation~\ref{eq:convective}, and recall that $\nabla_j y \propto \nabla_j (\epsilon-\delta)\propto -\nabla_j \delta$).  We further set $\nabla_i\delta\, \zeta_{ij}^{-1}\nabla_j\delta \approx 1.5\sigma_{11}^2(1-\gamma^2_{\nabla_\delta \nabla_\epsilon})/(x/3)$, where the numerator is the conditional variance of $\nabla\delta$ given that $\nabla\epsilon=0$, we approximated $\zeta_{ij}\approx\delta_{ij}x/3$ (retaining only the trace part), and 1.5 is a fudge factor to account for a similar term in equation \eqref{eq:dqdlnR}.} 
\begin{equation}
  \frac{\mathrm{d}y_R}{\mathrm{d} \ln R}
  =  -7 \left[\!\bigg(1+\frac{\mathrm{d}\ln\sigma_{23}}{\mathrm{d}\ln R^7}\bigg)\, y_R
    - \frac{\sigma_{22}}{\sigma_{23}}\,x_R \right]
\end{equation}
from equations \eqref{eq:yepsdel}, \eqref{eq:slopeEps} and \eqref{eq:x-slope}, and 
\begin{equation}
\label{eq:dqdlnR}
   \frac{\mathrm{d}q_R}{\mathrm{d} \ln R} = q_R\,\frac{\mathrm{d}\ln \sigma_{02}}{{\mathrm d}\ln R}.  
\end{equation}

\begin{figure}
 \centering
 \includegraphics[width=\columnwidth]{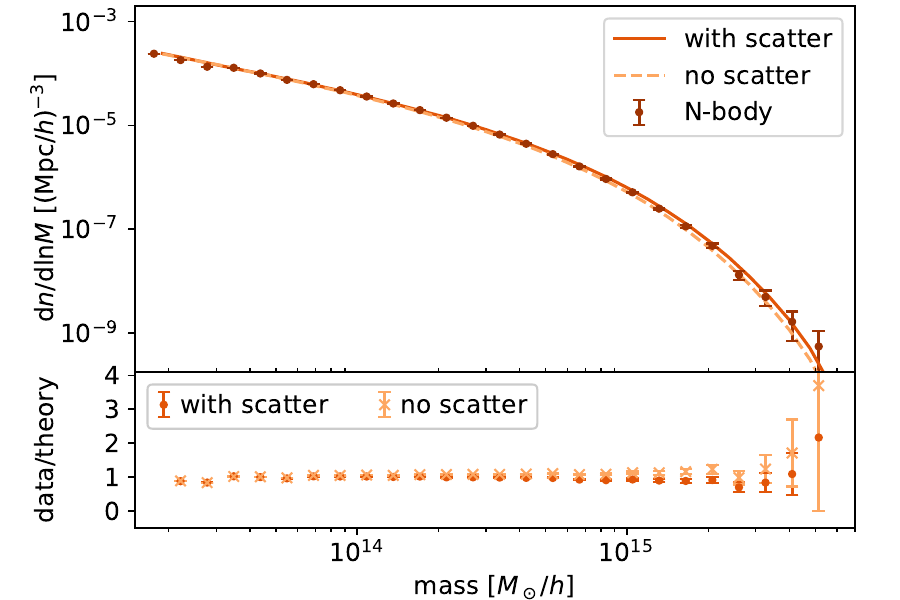}
 \caption{Comoving number density of haloes measured in simulations (error bars) and predicted by our energy-based approach (equation~\ref{eq:expksq}) which accounts for the correlation shown in Figure~\ref{fig:ey-sig}. Solid curve includes scatter around this correlation as described in the text, and dashed curve assumes there is none.  Bottom panel shows the fractional difference between the predictions and measurements.}
 \label{fig:dndlnm}
\end{figure}

Equation \eqref{eq:ycscatter} shows that assuming $\epsilon_c(y,q)$ brings additional dependence on $x$ and $y$ (as well as $q$) into $y_c$, which affects the predicted abundances as well as mean and variance of the typical overdensity of protohalo patches (equations \ref{eq:dpk} and \ref{eq:vpk}).
Some integrals in this example actually have analytical expressions. However, we do not give them here as the main point is simply to illustrate that our energy-peaks approach can be extended in a relatively transparent manner to include more involved collapse criteria.  Exploring what these criteria are is work in progress.  To motivate this exploration, Figure~\ref{fig:dndlnm} shows that equations \eqref{eq:expksq} and \eqref{eq:ecyq}, with $(a_0,a_1,a_2)=(1.75,0.21,0.42)$, provides a good description of halo abundances in our simulation.  

\section{Discussion and conclusions}
The assumption that haloes form from the spherical collapse of a homogeneous sphere has motivated the development of models in which haloes are identified with sufficiently dense peaks in the initial overdensity fluctuation field.  For a homogeneous sphere, the enclosed overdensity (equation~\ref{eq:d}) and energy (equation~\ref{eq:E}) are the same.  However, if the sphere is not homogeneous -- either because of substructure within it, or simply because its radial density profile, while smooth, is not flat -- then they are not.

We argued that a sphere in a realistic (non-spherical) density field follows spherical collapse as closely as possible if centered at a minimum of the enclosed total energy, rather than maximum of the mean matter density. At such minima, the center of convergence of the gravitational flow and its geometrical center coincide (equation~\ref{eq:dipole}).  We showed how to modify the excursion set peaks approach if haloes are identified with peaks in energy (equation \ref{eq:We}) rather than enclosed density (equation~\ref{eq:Wth}).  In this respect, it is natural to call ours the $\epsilon$ESP approach, and the traditional approach $\delta$ESP.  However, to ease notation, we simply used ESP throughout.

Our analysis and result (equation~\ref{eq:expks}) are no more complicated than the usual overdensity-based analysis.  The model naturally predicts scatter in the matter overdensity $\delta_R$ of protohalo patches (equation~\ref{eq:vpk}) around a mean value.  This mean decreases with $R$ (i.e., at lower masses) (equation~\ref{eq:dpk} and Figure~\ref{fig:conddelta}) if protohalo patches all have the same critical value of the energy overdensity $\epsilon_c$ whatever their size $R$.  However, both N-body simulations and theoretical arguments suggest that $\delta_R$ must be larger at smaller masses. Therefore, our model predicts that $\epsilon_c$ must increase even more strongly as protohalo patch size decreases.

Measurements in simulations confirm many of the generic predictions of our $\epsilon$ESP approach.  Maps of $\delta_R$ and $\epsilon_R$ show that protohalo patches are much more likely to be centered on $\epsilon_R$- than $\delta_R$-peaks (Figures~\ref{fig:eps}--\ref{fig:cumulative}).  The excursion set prediction that $\epsilon_R\ge \delta_R$ (equation~\ref{eq:yepsdel}) is also confirmed (Figure~\ref{fig:diff}), as is the prediction that $\epsilon_R$ should be a steeper function of mass than is $\delta_R$ (Figure~\ref{fig:epsc} and equations~\ref{eq:fitd} and~\ref{eq:fite}).
Although we only tested our formalism on halos more massive than $10^{13} h^{-1} M_\odot$, and find that it works better at higher masses, our results strongly suggest that $\epsilon_R$ is much better than $\delta_R$ for modeling smaller masses as well.
Our $\epsilon$ESP approach provides an excellent description of the speeds of protohalo patches (equations~\ref{eq:rmsvpk} and Figure~\ref{fig:pdfv}).  It also provides a good description of halo abundances (Figure~\ref{fig:dndlnm}) provided we include the fact that there is some stochasticity in $\epsilon_R$ values, even at fixed mass (Figure~\ref{fig:ey-sig} and equations~\ref{eq:ecyq} and~\ref{eq:expksq}).  
This motivates further study of our energy-based excursion set peaks model.

\begin{itemize}
\item  An obvious direction for future work is to check if energy- rather than density-based halo finders are less stochastic in the late time field. 
\item A more ambitious goal is to determine $\epsilon_c$ from first principles, including the mean trends and stochasticity shown in Figures~\ref{fig:epsc} and~\ref{fig:ey-sig}, and how these depend on the algorithm used to identify haloes in the simulations. This necessarily requires modeling the virialization process (e.g., using energy conservation arguments).
\item A closely related effort is to extend the analysis to allow triaxiality.  This ingredient is required because some (but not all!) of the stochasticity in protohalo matter overdensity correlates with the initial tidal shear field \citep{smt01,HahnPorcDekCar08,borzyShear}.  In models of triaxial collapse \cite[e.g.][]{bm96,monaco97,ab10,porciani11}, the leading order departure from sphericity is quadrupolar.  In this respect, our energy-based approach is attractive because choosing locations that are stationary points of the energy sets the dipole to zero but leaves all other multipoles unconstrained.  I.e., $\epsilon$ESP provides a natural setup for including deviations from sphericity. 
These effects would add to the stochasticity in $\delta_R$ discussed in this paper, which exists even in a purely spherical configuration.
\item Another direction is to estimate the implications for the spatial distribution of the protohalo patches: halo `bias'.  An excursion set-based model for halo abundances (like any analytical model) carries with it a description of halo bias \citep{st99,mps12,cps17,ModiCastSel16}. However, protohalo patches appear to be better described by a smoothing window that oscillates less strongly than $W_{\rm TH}=W_1$ \citep{css17}; our $W_{2}$ (equation~\ref{eq:We}) is indeed more damped than $W_{\rm TH}$ (equation~\ref{eq:Wth}), so it has qualitatively the right behaviour.  Moreover, $W_{2}$ has zeroes on slightly larger scales $x=kR$ than $W_{\rm TH}$ (Figure~\ref{fig:filters}), which is also qualitatively consistent with the measurements.  This, with the modified velocity statistics and associated `velocity bias' of equations~(\ref{eq:rmsvpk}), provides an additional test of our approach.
\item Excursion sets also provide a natural framework for describing the phenomenon known as assembly bias:  haloes of fixed mass may cluster differently depending on their assembly history \citep{st04,gsw05,fw10,lms17,borzyShear}.  Models based on a smoothing filter that is a tophat in $k$-space (rather than $W_1$), have $\delta_R$ that is Markov: in effect, the memory of larger scales is absent in such models.  For Markov models to exhibit assembly bias, other variables than $\delta_R$, such as the shear, must matter for halo formation \citep{kn07,cs13}.  For density-based models that use a Top-Hat filter $W_1$, only $\delta_R$ and its slope can be used to explain assembly bias phenomena, because statistical moments with more powers of $k$ (like the peak curvature or the second $R$-derivative) diverge \citep{markov14}, effectively making the slope of $\delta_R$ Markov.  Gaussian smoothing allows density and second derivatives to be used \citep{zentner06,dwbs08}, but because slope and curvature are the same for this filter, there is, in effect, only one additional parameter that can be used to model assembly bias phenomena.  Moreover, the connection to the physics of spherical collapse is lost.  In our energy-based approach -- i.e. smoothing with $W_2$ -- the second derivatives are statistically well behaved, and slope and curvature are distinct.  This potentially allows physically realistic models of additional halo properties.
\item The excursion set troughs model uses an overdensity-based approach to study abundance and clustering of voids in the large scale matter distribution \citep{svdw04,pls12voids,JenningsVoids13,AchiNeyPar13,ms19}.  It is natural to study if our energy-based approach is useful for voids and other constituents of the cosmic web.  If so, incorporating it into the `skeleton' framework \citep{h01,spcnp08,codisSpin,bweb18} or other approaches \citep{bkp96,sams06,tetra16}, should be as straightforward as it was for peaks.
\item
Finally, our approach is relevant in the context of modeling primordial black hole abundances \citep{nakamaPBH,gm19} because there too, the condition for formation involves multiple variables for which the total Jacobian is no longer the simple product of the peak determinant times the upcrossing term, so the extra complications treated in Appendix~\ref{app:alt} are necessary \citep{gs20,ymPBH20}.
\end{itemize}

\section*{Acknowledgements}
We are grateful to the ICTP for its hospitality over the years, but especially during the summer of 2014 when we first discussed the main idea presented here, and to the IFPU and the Munich Institute for Astro- and Particle Physics (MIAPP) of the DFG cluster of excellence ``Origin and Structure of the Universe'' for their hospitality during July 2019 when this manuscript was completed.  We thank Giulia Despali for sharing the data of the Flora simulation with us, and Corentin Cadiou for many helpful discussions on numerical aspects.

\section*{Data availability statement}
The data of the Flora simulation, and those underlying this article, can be shared on reasonable request to the corresponding author.

\bibliographystyle{mnras}
\bibliography{mybib} 

\appendix

\section{Multipole expansion}
\label{app:acc}

The Fourier modes of the rescaled peculiar acceleration $-\nabla\phi$ are $(i \bm{k}/k^2)\delta({\bm k}) \mathrm{e}^{i\bm{k}\cdot\bm{r}}$. Those of its average over $V$ have $\mathrm{e}^{i\bm{k}\cdot\bm{r}}$ replaced by $3j_1(kR)/kR$. Those of $\bm{r}_{\mathrm{cm}}/3 = \bm{D}_RR/3$ have it replaced by $j_2(kR)$.
Since $j_2(x)-3j_1(x)/x = -j_0(x)$, equation \eqref{eq:g} gives
\begin{equation}
  \frac{\bm{g}}{4\pi G\bar\rho} = \int \frac{{\rm d}{\bm k}}{(2\pi)^3}
  \frac{i \bm{k}}{k^2}\delta({\bm k})
  \bigg[\mathrm{e}^{i\bm{k}\cdot\bm{r}}-j_0(kR) \bigg]
  - \frac{\bm{r}}{3} \,.
\label{eq:gm}
\end{equation}

Expanding the plane wave in spherical harmonics  returns 
\begin{equation}
  \mathrm{e}^{i\bm{k}\cdot\bm{r}} = \sum_{l=0}^\infty \frac{i^l}{l!} Y_{i_1\dots i_l}(\bm{\hat k})Y_{i_1\dots i_l}(\bm{\hat r})j_l(kr)(2l+1)!!
\end{equation}
where $ Y_{i_1\dots i_l}(\bm{\hat r}) = [r^{l+1}/(2l-1)!!]\pd^l(1/r)/(\pd r_{i_1}\dots\pd r_{i_l})$ is a modified version of the real spherical harmonics. Their indices are by construction totally symmetric and traceless.
The first few of them are $Y=1$, $Y_i=\hat r_i$, $Y_{ij}=\hat r_i\hat r_j-\delta_{ij}/3$ and $Y_{ijk}=\hat r_i\hat r_j\hat r_k-(\delta_{ij}\hat r_k+\delta_{ik}\hat r_j+\delta_{jk}\hat r_i)/5$.

For $r=R$, the $l=0$ term is just $j_0(kR)$, and drops out of equation \eqref{eq:gm}. The $l=1$ term gives $-\hat k_i\hat k_j \delta({\bm k}) W_1(kR) r_j$, and thus  $-[(\delta_{ij}/3)\delta_R + q_{ij}]r_j$, where $q_{ij}$ is the traceless shear. In general, the $Y$'s obey $\hat k_j Y_{i_1\dots i_l}(\bm{\hat k})Y_{i_1\dots i_l}(\bm{\hat r}) = Y_{ji_1\dots i_l}(\bm{\hat k})Y_{i_1\dots i_l}(\bm{\hat r}) + [l/(2l+1)]Y_{i_1\dots i_{l-1}}(\bm{\hat k})Y_{ji_1\dots i_{l-1}}(\bm{\hat r})$.
We can then write equation \eqref{eq:gm} as
\begin{align}
    \frac{\bm{g}}{4\pi G\bar\rho} &=-\frac{1+\delta_R}{3}r_i -q_{ij}r_j
  - D_j \bigg(\hat r_i\hat r_j-\frac{\delta_{ij}}{3}\bigg)R
  \notag \\
  &- \sum_{l=2}^\infty \frac{R}{l!}
  \bigg[q_{ji_1\dots i_{l}} Y_{i_1\dots i_{l}}(\bm{\hat r})
  + D_{i_1\dots i_{l}}Y_{ji_1\dots i_{l}}(\bm{\hat r})\bigg],
\end{align}
where
\begin{align}
  &D_{i_1\dots i_{l}} \equiv \int \frac{{\rm d}{\bm k}}{(2\pi)^3}
  i^lY_{i_1\dots i_{l}}\!(\bm{\hat k})\,\delta({\bm k})\,
  (2l+1)!!\frac{j_{l+1}(kR)}{kR} \,,\\
  &q_{i_1\dots i_{l}} \equiv -\!\int \!\frac{{\rm d}{\bm k}}{(2\pi)^3}
  \,i^lY_{i_1\dots i_{l}}\!(\bm{\hat k})\,\delta({\bm k})\,
  (2l-1)!!\frac{j_{l-1}(kR)}{kR} \,.
\end{align}
Multiplying by $4\pi G \bar\rho$ gives back equation \eqref{eq:gmult}, with $M=(4\pi R^3/3)\bar\rho(1+\delta_R)$, where $\delta_R$ is actually needed only in the first term and is of second order elsewhere.

\section{Correlations between variables}
\label{app:corrs}

In what follows, we will drop the explicit dependence on $R$.  
We compute the cross-correlation coefficients of the normalized variables $\omega=\epsilon/\sigma_{02}$, $\nu=\delta/\sigma_{01}$, $x$ and $y$ defined in Section \ref{sec:norm}. Since
\begin{align}
  & -\langle\epsilon\nabla^2\epsilon\rangle=\langle \nabla\epsilon\cdot\nabla\epsilon\rangle = \sigma_{12}^2\,\\
  & 2\langle\epsilon\mathrm{d}\epsilon/\mathrm{d} R\rangle = \mathrm{d}\langle\epsilon^2\rangle/\mathrm{d}R = \mathrm{d}\sigma_{02}^2/\mathrm{d}R \,\\
  & 2\langle\delta\mathrm{d}\delta/\mathrm{d}R\rangle = \mathrm{d}\sigma_{01}^2/\mathrm{d}R \,\\
  & -2\langle \nabla^2\epsilon\, (\mathrm{d}\epsilon /\mathrm{d}R)\rangle = (\mathrm{d}/\mathrm{d}R)\langle \nabla\epsilon\cdot\nabla\epsilon\rangle
  = \mathrm{d}\sigma_{12}^2/\mathrm{d}R\,\\
  & \langle\eta\cdot\nabla\delta\rangle = -
  \langle\delta\nabla^2\epsilon\rangle/\sigma_{12}
\end{align}
then in the same order
\begin{align}
  \gamma_{x\omega}&\equiv \langle x \omega\rangle
  = \frac{\sigma_{12}^2}{\sigma_{22}\sigma_{02}} \,, \\
  \gamma_{y\omega} &\equiv \langle y\omega\rangle
  = -\frac{7}{2}\frac{{\rm d}\sigma_{02}^2/\mathrm{d}\ln R}{R^2\sigma_{23}\sigma_{02}}
  = -\frac{7\sigma_{02}}{R^2\sigma_{23}}
  \frac{{\rm d}\ln\sigma_{02}}{\mathrm{d}\ln R}\,,\\
\label{eq:gammayom}
  \gamma_{x\nu} &\equiv \langle x\nu\rangle
  = -\frac{5}{2}\frac{{\rm d}\sigma_{01}^2/\mathrm{d}\ln R}{R^2\sigma_{22}\sigma_{01}}
  = -\frac{5\sigma_{01}}{R^2\sigma_{22}}
  \frac{{\rm d}\ln\sigma_{01}}{\mathrm{d}\ln R}\,,\\
  \gamma_{y x} &\equiv \langle yx\rangle
  = -\frac{7}{2}\frac{{\rm d}\sigma_{12}^2/{\rm d}\ln R}{R^2\sigma_{23}\sigma_{22}}
  =-\frac{7\sigma_{02}\gamma_{x\omega}}{R^2\sigma_{23}}\frac{{\rm d}\ln\sigma_{12}}{{\rm d}\ln R} \,,
\label{eq:gammaxyApp} \\
  \gamma_{\rm grad} &\equiv
  \frac{\langle\bm{\eta}\cdot\bm{\nabla}\delta\rangle}{\sigma_{11}}
  =  \frac{\sigma_{01}\sigma_{22}}{\sigma_{11}\sigma_{12}} \gamma_{x\nu}\,.
\end{align}
Notice that equations \eqref{eq:gammayom} and \eqref{eq:gammaxyApp} also imply that
\begin{equation}
  \frac{{\rm d}\ln\sigma_{12}}{{\rm d}\ln\sigma_{02}}
  = \frac{\gamma_{xy}}{\gamma_{x\omega}\gamma_{y\omega}}\,.
\end{equation}
Equation \eqref{eq:yepsdel} implies $\mathrm{d}\epsilon/\mathrm{d}\ln R=-(R^2\sigma_{23}/7)y=5(\delta-\epsilon)$. Therefore
\begin{equation}
  \frac{1}{10}\frac{{\rm d}\sigma_{02}^2}{\mathrm{d}\ln R}
  = \langle\delta\epsilon\rangle - \langle \epsilon\epsilon\rangle,
\end{equation}
so
\begin{equation}
  \langle\delta \epsilon\rangle = \langle \epsilon\epsilon\rangle\,\left[1 + \frac{1}{10}\frac{{\rm d}\ln\sigma_{02}^2}{{\rm d}\ln R}\right].
\end{equation}
The final term in square brackets is a measure of the stochasticity between $\delta_R$ and $\epsilon_R$.  The normalized correlation coefficient between the two is
\begin{equation}
  \gamma_{\nu\omega} = \langle\nu\omega\rangle
  = \frac{\sigma_{02}}{\sigma_{01}}\,
  \left[1 + \frac{1}{10}\frac{{\rm d}\ln\sigma_{02}^2}{{\rm d}\ln R}\right];
\end{equation}
this is closer to unity when the large scale power dominates over that from smaller scales.  
Similarly,
\begin{equation}
  \gamma_{y x} =
  \frac{35}{R^2\sigma_{23}}(\langle\epsilon x\rangle - \langle\delta x\rangle),
\end{equation}
which is therefore equivalent to equation \eqref{eq:gammaxy}.

\begin{figure}
  \includegraphics[width=\columnwidth]{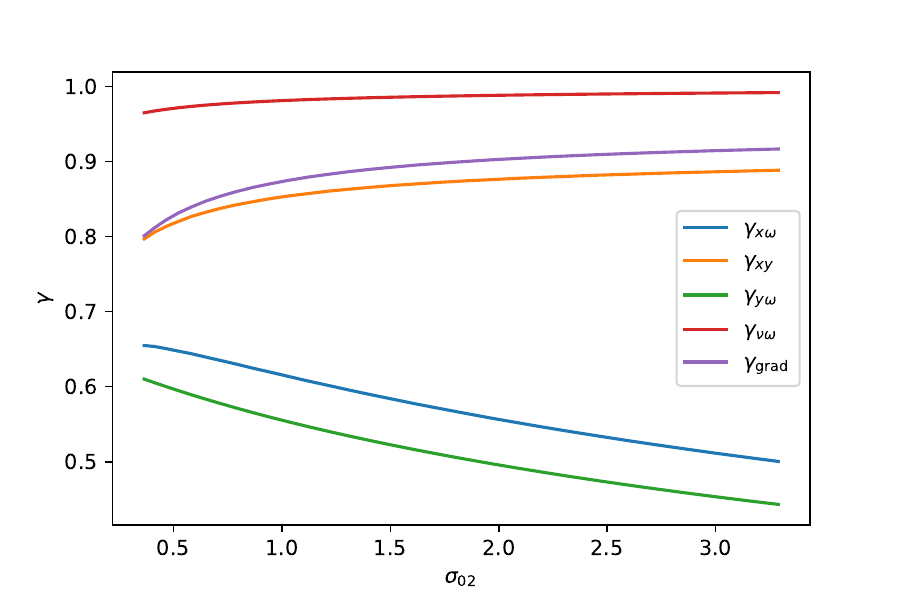}
  \caption{\label{fig:gammas} Behavior of various cross-correlation coefficients $\gamma_{ab}$ (as labeled) for a $\Lambda$CDM power spectrum:  $\gamma_{\nu\omega}\approx 0.95$ indicates that energy and overdensity are very strongly correlated; $\gamma_{\rm grad}\approx 0.85$ indicates that while many energy peaks will be overdensity-peaks, there will certainly be some peaks in energy which are not peaks in overdensity, and vice-versa (because $\gamma_{\rm grad}\ne 1$). }
\end{figure}

In addition, since $\langle(\mathrm{d}\epsilon/\mathrm{d}\ln R)^2\rangle=(R^2\sigma_{23}/7)^2$, then
\begin{align}
  \bigg(\frac{R^2\sigma_{23}}{35}\bigg)^2
  &= [\sigma_{01}^2 + \sigma_{02}^2 - 2\langle\delta \epsilon\rangle] \notag\\
  &= \sigma_{02}^2 \left[\frac{\sigma_{01}^2}{\sigma_{02}^2} - 1 - \frac{{\rm d}\ln\sigma_{02}^2}{{\rm d}\ln R^5}\right].
\end{align}
For power-law spectra, the term in square brackets is a constant which decreases as the slope of $P(k)$ decreases (becomes more negative).  
Equations \eqref{eq:gammayom} and \eqref{eq:gammaxy} then give
\begin{align}
  \gamma_{y x} 
  &=-\frac{\gamma_{x\omega}}{2}\,\frac{{\rm d}\ln\sigma_{12}^2/{\rm d}\ln R^5}{\sqrt{\sigma_{01}^2/\sigma_{02}^2 - 1 - {\rm d}\ln\sigma_{02}^2/{\rm d}\ln R^5}}\,,\\
  \gamma_{y\omega}
  & = -\frac{1}{2}\frac{{\rm d}\ln\sigma_{02}^2/{\rm d}\ln R^5}{\sqrt{\sigma_{01}^2/\sigma_{02}^2 - 1 - {\rm d}\ln\sigma_{02}^2/{\rm d}\ln R^5}}\,.
\end{align}

Figure~\ref{fig:gammas} shows how a number of these cross-correlation coefficients for a $\Lambda$CDM power spectrum depend on smoothing scale $R$, parametrized by $\sigma_{02}$.  To understand this scale dependence, it is useful to study the case in which $P(k) \propto k^n$.  Then ${\rm d}\ln\sigma_{jl}^2/{\rm d}\ln R = -(2j+n+3)$ (independently of the filter), and, for $n=(0,-1,-2)$, 
\begin{align}
 (\sigma_{01}/\sigma_{02})^2 &= (7/10, 18/25, 21/25),\\
  (\sigma_{02}/R\sigma_{12})^2 &= (2/21,1/6,1/3)\\
  (\sigma_{12}/R\sigma_{22})^2 &= (\mathrm{ND},\mathrm{ND},2/21)\\
  \gamma_{x\omega} &= (\mathrm{ND},\mathrm{ND},\sqrt{2/7}) \\
  \gamma_{y\omega} &= (\sqrt{3/10},1/\sqrt{3},1/2) \\
  \gamma_{yx} &= (\mathrm{ND},\mathrm{ND},3/\sqrt{14}) \\
   \gamma_{\nu\omega} &= (\sqrt{7/10}, \sqrt{8/9}, \sqrt{27/28}),
\end{align}
where ND indicates that the coefficient contains divergent integrals.

\section{A threshold for $\delta_R$}
\label{app:alt}

We discuss here how our formalism should be modified if one wanted to introduce a critical value for the density $\delta_R$, while still fixing the position $\bm{x}$ through the constraint $\nabla\epsilon_R=0$.
In this case, using the formalism introduced in Section \ref{sec:formal}, the number density of stationary points would be
\begin{equation}
  |\tilde J|\,\delta_{\rm D}^{(3)}(\bm{\eta})\,
  \delta_{\rm D}(\delta_R-\delta_c)
\end{equation}
where $\tilde J = \det[\partial\{\delta_R-\delta_c,\bm{\eta}\}/\partial\{R,\mathbf{x}\}]$ is the 4-dimensional Jacobian determinant. However, now $\nabla_i(\delta_R-\delta_c)\neq0$, since the peak constraint is enforced on $\epsilon_R$  and not on $\delta_R$, and $\tilde J$ no longer factorizes in a simple way. Assuming for simplicity that $\nabla_i\delta_c=0$, one has
\begin{equation}
  |\tilde J| = |\det(\nabla\bm{\eta})|
  \bigg|\frac{\pd\delta}{\pd R}
  -\frac{\pd\eta_i}{\pd R}(\nabla_i\eta_j)^{-1}\nabla_j\delta\bigg|\,.
\end{equation}
The total determinant is no longer the simple product of the peak determinant times the upcrossing term $\pd\delta/\pd R$.

We recall that the position $\bm{x}_{\mathrm{pk}}$ of the peak (or in general of the stationary point) depends on the smoothing scale $R$.  Therefore, we introduce the ``convective'' derivative
\begin{equation}
  \label{eq:operator}
  \frac{\mathrm{d}}{\mathrm{d}R} \equiv \frac{\pd}{\pd R} +
  \frac{\mathrm{d}\bm{x}_{\mathrm{pk}}}{\mathrm{d}R}\cdot\nabla
\end{equation}
which accounts for this displacement. In order to preserve the constraint $\eta_i=0$ across scales, one must impose that $\mathrm{d}\eta_i/\mathrm{d}R = 0$, which implies that
\begin{equation}
  \frac{\mathrm{d}x_{j,\mathrm{pk}}}{\mathrm{d}R} = -
  \frac{\pd\eta_i}{\pd R}(\nabla_i\eta_j)^{-1}\,.
\end{equation}
Hence, the Jacobian determinant can be written as
\begin{equation}
  |\tilde J| = |\det(\nabla\bm{\eta})|
  \bigg|\frac{\mathrm{d}\delta}{\mathrm{d} R}\bigg|
  = \bigg(\frac{\sigma_{22}}{\sigma_{12}}\bigg)^3|\det(\zeta)|
  \bigg|\frac{\mathrm{d}\delta}{\mathrm{d} R}\bigg|\,,
\end{equation}
which describes the crossing along the trajectory that follows the stationary point of $\epsilon_R$.
Moreover, since $\eta_i = \nabla_i\epsilon/\sigma_{12}$ and $\pd W_2/\pd R = (5/R)(W_1 - W_2)$, we also have that
\begin{equation}
  \frac{\pd\eta_i}{\pd R} =
  -\bigg(\frac{5}{R}+\frac{\pd\ln\sigma_{12}}{\pd R}\bigg) \eta_i
  + \frac{5}{R}\frac{\nabla_i\delta}{\sigma_{12}}\,,
\end{equation}
where the first term vanishes because $\eta_i=0$. Therefore, equation~(\ref{eq:operator}) becomes 
\begin{equation}
  \frac{\mathrm{d}}{\mathrm{d} R}
  = \frac{\pd}{\pd R} + \frac{5}{R}
  \frac{\nabla_i\delta\,\zeta_{ij}^{-1}\nabla_j}{\sigma_{22}}\,,
\label{eq:convective}
\end{equation}
where we have used equations~\eqref{eq:etaR} and~\eqref{eq:zetaR} to set $\sigma_{12}\nabla _i\eta_j=\sigma_{22}\,\zeta_{ij}$.  Equation~\eqref{eq:convective} shows that if $\bm{x}_{\mathrm{pk}}$ (a stationary point of $\epsilon_R$) is also a stationary point of $\delta_R$, i.e. if both $\bm{\nabla}\epsilon_R=0$ and $\bm{\nabla}\delta_R=0$, then convective and partial derivatives of {\em any} field are equal.

Since $\zeta_{ij}$ is positive definite due to the peak constraint, so is $\zeta_{ij}^{-1}$, and the second term from the above equation applied to $\delta$ is always positive.  It can then happen that $\mathrm{d}\delta/\mathrm{d} R>0$ even if $\pd\delta/\pd R<0$. Hence, requesting upcrossing along the peak trajectory imposes the stricter constraint
\begin{equation}
  x_R> \bigg(\frac{5\sigma_{11}}{R\sigma_{22}}\bigg)^2
  \frac{\nabla_i\delta\,\zeta_{ij}^{-1}\nabla_j\delta}{\sigma_{11}^2}\,,
\end{equation}
with $x_R$ as in Equation \eqref{eq:x-slope}.
If $\lambda_1>\lambda_2>\lambda_3>0$ are the eigenvalues of $\zeta_{ij}$, the inequality above is equal to
\begin{equation}
  \lambda_1+\lambda_2+\lambda_3 > \bigg(\frac{5}{R\sigma_{22}}\bigg)^2
  \bigg[\frac{(\nabla_x\delta)^2}{\lambda_1}+
  \frac{(\nabla_y\delta)^2}{\lambda_2} +
  \frac{(\nabla_z\delta)^2}{\lambda_3}\bigg],
\end{equation}
where we denoted $x$, $y$ and $z$ the Cartesian coordinates in the frame in which $\zeta_{ij}$ is diagonal. This constraint reduces to the ordinary upcrossing condition $x_R>0$ if $\bm{\nabla}\delta=0$.

\bsp	
\label{lastpage}
\end{document}